\documentclass[fleqn,usenatbib]{mnras}
\usepackage[T1]{fontenc}
\usepackage{graphicx}	
\usepackage{amsmath}
\usepackage{amssymb}
\usepackage{ae,aecompl}
\usepackage{txfonts}
\usepackage{colortbl}

\usepackage{soul}
\usepackage{comment}
\usepackage{ulem} 

\DeclareRobustCommand{\VAN}[3]{#2}
\let\VANthebibliography\thebibliography
\def\thebibliography{\DeclareRobustCommand{\VAN}[3]{##3}\VANthebibliography}

\begin{document}
\title[AGN are important in dwarfs]{
Radio AGN in nearby dwarf galaxies: the important role of AGN in dwarf-galaxy evolution}

\author[F. Davis et al.]{F. Davis$^{1}$,
S. Kaviraj$^{1}$\thanks{E-mail: s.kaviraj@herts.ac.uk}, M. J. Hardcastle$^{1}$, G. Martin$^{2,3}$, R. A. Jackson$^{4}$, K. Kraljic$^{5}$, K. Malek$^{6, 7}$, \newauthor  S. Peirani$^{8,9}$, D. J. B. Smith$^{1}$, M. Volonteri$^{10}$ and L. Wang$^{11,12}$
\\
$^{1}$Centre for Astrophysics Research, Department of Physics, Astronomy and Mathematics, University of Hertfordshire, Hatfield, AL10 9AB, UK\\
$^{2}$Korea Astronomy and Space Science Institute, 776 Daedeokdae-ro, Yuseong-gu, Daejeon 34055, Korea\\
$^{3}$Steward Observatory, University of Arizona, 933 N. Cherry Ave, Tucson, AZ 85719, USA\\
$^{4}$Department of Astronomy and Yonsei University Observatory, Yonsei University, Seoul 03722, Republic of Korea\\
$^{5}$Aix Marseille Univ, CNRS, CNES, LAM, Marseille, France\\
$^{6}$National Centre for Nuclear Research, ul.Pasteura 7, 02-093 Warszawa, Poland\\
$^{7}$Aix Marseille Univ. CNRS, CNES, LAM Marseille, France\\
$^{8}$Université Cote d’Azur, Observatoire de la Cote d’Azur, CNRS, Laboratoire Lagrange, Bd de l’Observatoire,
CS 34229, 06304 Nice Cedex 4, France\\
$^{9}$Institut d’Astrophysique de Paris (UMR 7095: CNRS \& UPMC), 98 bis Bd Arago, 75014 Paris, France\\
$^{10}$Institut d’Astrophysique de Paris, Sorbonne Université, CNRS, UMR 7095, 98 bis bd Arago, 75014 Paris, France\\
$^{11}$SRON Netherlands Institute for Space Research, Landleven 12, 9747 AD, Groningen, The Netherlands\\
$^{12}$Kapteyn Astronomical Institute, University of Groningen, Postbus 800, 9700 AV Groningen, the Netherlands}

\label{firstpage}
\pagerange{\pageref{firstpage}--\pageref{lastpage}}
\maketitle

\begin{abstract}
We combine deep optical and radio data, from the Hyper Suprime-Cam and
the Low-Frequency Array (LOFAR) respectively, to study 78 radio AGN in
nearby ($z<0.5$) dwarf galaxies. Comparison to a control sample,
matched in stellar mass and redshift, indicates that the AGN and
controls reside in similar environments, show similar star-formation
rates (which trace gas availability) and exhibit a
comparable incidence of tidal features (which indicate recent
interactions). We explore the AGN properties by combining the
predicted gas conditions in dwarfs from a cosmological hydrodynamical
simulation with a Monte-Carlo suite of simulated radio sources, based
on a semi-analytical model for radio-galaxy evolution. In the subset
of LOFAR-detectable simulated sources, which have a similar
distribution of radio luminosities as our observed AGN, the median jet
powers, ages and accretion rates are $\sim 10^{35}$ W, $\sim 5$ Myr
and $\sim 10^{-3.4}$ M$_{\odot}$ yr$^{-1}$ respectively. The median mechanical energy output of these sources is $\sim 100$ times larger than the median binding energy expected in dwarf gas reservoirs, making AGN feedback plausible. Since special circumstances (in terms of environment, gas availability and interactions) are not necessary for the presence of AGN, and the central gas masses are predicted to be an order of magnitude larger than that required to fuel the AGN, AGN triggering in dwarfs is likely to be stochastic and a common phenomenon. Together with the plausibility of energetic feedback, this suggests that AGN could be important drivers of dwarf-galaxy evolution, as is the case in massive galaxies. 
\end{abstract}



\section{Introduction}
\label{sec:Intro}

Our current statistical understanding of galaxy evolution is largely underpinned by massive galaxies. This is primarily because these systems are bright enough to be detectable, across cosmic time, in past wide-area surveys, which offer large footprints but are relatively shallow. However, while massive galaxies form the basis of most of the existing literature, their dwarf (M$_*$ < 10$^{9.5}$ M$_{\odot}$) counterparts dominate the galaxy number density in all environments and at all epochs \citep[e.g.][]{Wright2017,Martin2019}. A complete understanding of galaxy evolution therefore demands a comprehension of how dwarf galaxies form and evolve over time \citep[e.g.][]{Calabro2017}.

The dearth of statistical studies in the dwarf regime, particularly outside the very local Universe, is driven primarily by the fact that \textit{typical} dwarfs are too faint to be visible in past wide-area surveys like the Sloan Digital Sky Survey (SDSS, \citealt{Alam2015}). The dwarfs that are visible in such surveys tend to be anomalously bright, due to the presence of high levels of star formation, driven by recent mergers or fly-bys with massive galaxies \cite[e.g.][]{Jackson2021a}. This makes these systems transient and unrepresentative of the general dwarf population, making it difficult to draw conclusions about how typical dwarfs evolve from such surveys. 

Our past inability to study statistical samples of typical dwarfs over cosmic time has two important implications. First, our empirical picture is biased towards bright (massive) galaxies. Second, since our current galaxy formation models are calibrated to reproduce only the subset of massive galaxies, our understanding of the physics of galaxy evolution is potentially highly incomplete. Not unexpectedly, many well-known inconsistencies between theory and observation are found in the dwarf regime, e.g. the classical `substructure' \citep{Moore1999,Bullock2017}, `core-cusp' \citep{Blok2010}, `too-big-to-fail' \citep{Boylan-Kolchin2011} and `satellite planes' problems \citep[e.g.][]{Kroupa2005,Metz2007,Tully2015}, or more recent controversies, such as the presence of severely dark-matter-deficient dwarf galaxies \citep[e.g.][but see \citealt{Jackson2021b} for a solution to this problem]{vanDokkum2018,Guo2019}.

The advent of new optical surveys that are both deep and wide, like the recent Hyper Suprime-Cam Subaru Strategic Program \citep[HSC-SSP,][]{Aihara2018a} and the forthcoming Legacy Survey of Space and Time \citep[LSST, e.g.][]{Robertson2017}, enables us to perform some of the first unbiased, statistical studies of the general dwarf galaxy population, out to at least intermediate redshift ($z\sim0.5$). When combined with deep-wide surveys at other wavelengths, such as the radio (as is the case here), this new generation of datasets offers an unprecedented opportunity to study many aspects of galaxy evolution in the dwarf regime that we were previously restricted to exploring only in massive galaxies.

Active galactic nuclei (AGN) are thought to play an important role in regulating the evolution of massive galaxies \citep[e.g.][]{Kaviraj2007,Schawinski2007,Hardcastle2007,Pipino2009,Shabala2011,Fabian2012}, with negative AGN feedback routinely employed in galaxy formation models to bring the predicted properties -- e.g. stellar masses, morphologies, colours and star formation rates (SFRs) -- of massive galaxies in line with observations \citep[e.g.][]{Cole2000,Bower2006,Croton2006,Beckmann2017,Kaviraj2017}. However, whether black holes (BHs) exist in dwarf galaxies in appreciable numbers, and how these BHs may influence dwarf-galaxy evolution (e.g. through AGN feedback), remain important but largely unexplored questions, both from empirical and theoretical standpoints \citep[e.g.][]{Volonteri2008,VanWassenhove2010,Greene2020}.

The generic presence of BHs and AGN feedback in dwarf galaxies could offer solutions to many unresolved problems in our standard paradigm \citep[e.g.][]{Volonteri2010,Silk2017}. For example, AGN feedback could help suppress the excess number of dwarfs seen in simulations \citep[e.g.][]{Kaviraj2017} and bring dwarf galaxy number densities in line with observational data, which supernova feedback alone may fail to do \citep[e.g.][]{Keller2016}. Ejection of material through AGN feedback may provide a solution to the `over-massive' dwarf galaxies predicted by simulations \citep{Garrison-Kimmel2013}. It may also explain the large fraction ($\sim 30$ per cent) of baryons that is missing from massive discs, that must accumulate at least some of their mass through the accretion of dwarfs \citep{Shull2012}. Accreting intermediate mass BHs in dwarfs could contribute to re-ionization \citep[e.g][]{Volonteri2009} and provide natural seeds for the hierarchical formation of the supermassive BHs that are ubiquitously found in massive galaxies \citep[e.g.][]{Johnson2013}. 

While BHs in dwarf galaxies are included in current cosmological simulations, BH growth in this regime is stunted \citep[e.g.][]{Dubois2015,Bower2017,Habouzit2017,McAlpine2018,Trebitsch2018}, both due to the displacement of gas around the BH by stellar feedback, and because BHs tend to wander in the simulated dwarfs due to their shallow potential wells, and often spend very little time in regions of high gas density. However, it is important to note that current cosmological simulations do not have the spatial resolution to properly resolve the vicinity of the BH \citep[e.g.][]{Beckmann2018,Beckmann2019}. It is, therefore, not clear whether these effects are an artefact of the relatively low spatial resolution of such simulations or whether BHs in dwarfs in the real Universe may actually exhibit the same trends. Observational studies of AGN in dwarf galaxies, across cosmic time, are therefore a critical exercise, to put constraints on a key element of our structure-formation paradigm. 

Previous observational work on AGN in dwarfs has been hampered by the difficulties in detecting dwarf galaxies in past optical surveys (as described above), coupled with the challenges in detecting the low-mass black holes that reside in these systems. Nevertheless, a burgeoning literature has used the accretion signatures of active BHs to study their properties in dwarf galaxies using, for example, optical emission-line diagnostics \citep[][]{Greene2007,Reines2013,Moran2014}, optical variability \citep[e.g.][]{Baldassare2020b}, infrared photometry \citep{Jarrett2011,Satyapal2014,Marleau2017,Satyapal2018,Kaviraj2019}, nuclear X-ray emission \citep{Secrest2015,Mezcua2016,Pardo2016,Chen2017,Baldassare2017,Mezcua2018,Birchall2020} and excess radio emission that cannot be accounted for by star formation alone \citep{Nyland2017, Mezcua2018, Mezcua2019}. 

Several recent studies, that have harnessed such methods have identified large statistical samples of AGN in dwarf galaxies \citep[e.g.][]{Greene2007,Reines2013,Moran2014,Satyapal2014, Sartori2015,Lemons2015,Nucita2017,Marleau2017,Aird2018,Mezcua2018,Mezcua2019,Kaviraj2019,Greene2020,Baldassare2020b,Birchall2020}. These studies show that massive BHs can indeed form in dwarfs and can experience sustained periods of accretion resulting in AGN activity, with the AGN fractions ranging from fractions of a per cent to a few tens of per cent. Given the difficulty in detecting both dwarfs and their central BHs at cosmological distances, these are strict lower limits. Indeed, in very nearby objects ($cz<10,000$ km s$^{-1}$) where high S/N observations of dwarfs can be made, some studies indicate that the fraction of systems that exhibit AGN signatures can be high \citep[e.g. $\sim$80 per cent][]{Dickey2019} and a full BH occupation fraction cannot be ruled out \citep[e.g.][]{Miller2015}. 

Recent studies have increasingly focused on the properties and demographics of these BHs and their potential role in influencing the evolution of their dwarf galaxy hosts. For example, \citet{Birchall2020} have used X-ray observations to demonstrate that AGN in dwarfs can have a broad range of accretion rates. 
\citet{Kaviraj2019} have combined deep optical data from the HSC-SSP with infrared photometry from \textit{WISE} \citep{Wright2010} to show that AGN in dwarfs are likely triggered by secular processes (rather than mergers) and that their bolometric luminosities indicate that AGN feedback is plausible and is likely to be an important driver of the evolution of their host galaxies.  

Resolved optical emission-line measurements in dwarf galaxies have allowed studies of the nuclear regions of these systems in unprecedented detail. Many of these studies have found evidence of emission-line ratios that are indicative of AGN activity \citep[e.g.][]{Dickey2019,Mezcua2020}. In some cases, the spatially-resolved kinematics show evidence for AGN-driven outflows in low-redshift dwarfs  \citep[e.g][]{Penny2018,Manzano-King2019,Liu2020}. In a similar vein, recent radio studies \citep[e.g.][]{Mezcua2019} have shown that radio jets in dwarfs (out to $z \sim 3.4$) have comparable efficiencies, and are capable of producing the same type of mechanical feedback, as those found in massive galaxies. Indeed, it is worth noting that dwarfs lie on an extrapolation of the M - $\sigma$ relation traced by massive galaxies, with similar scatter \citep[e.g.][]{Schutte2019,Baldassare2020a,Greene2020,Davis2020}. This suggests that AGN activity could be important in this regime, because BH feedback predicts a universal M $\propto$ $\sigma^4$ relation \citep[e.g.][]{Silk1998,King2015,King2021}. The growing empirical evidence for AGN feedback appears consistent with the results of recent (idealised) high-resolution simulations, which have shown that energetic feedback from accreting BHs in dwarfs can indeed influence the SFRs in their host galaxies, in a similar way to what is seen in massive galaxies \citep[e.g.][]{Barai2019,Koudmani2019,Koudmani2021}. 

Most techniques used to identify AGN in dwarfs so far have some potential drawbacks. 
Detection of AGN via optical emission lines, in shallow surveys like the SDSS, may miss a significant percentage of the dwarf-AGN population, both due to the fact that many systems will not exhibit emission lines with adequate S/N to be measured, and also because the dwarfs that are detected by these surveys are, as noted above, heavily biased towards highly star-forming systems \citep[e.g.][]{Baldassare2020b, Jackson2021a}. Infrared-selected AGN may include some contamination from star-forming systems although, given the properties of their interstellar media, this is less likely at low redshift \citep[e.g.][]{Satyapal2018}. Identifying low-mass BHs via their X-ray emission can also be problematic, because separating the weak X-ray emission due to accreting low-mass BHs from potential contaminants like X-ray binaries and emission from the hot interstellar gas can be a challenge \citep[][]{Birchall2020}. 

Selecting AGN in dwarfs via their radio emission offers a route to
efficiently and unambiguously confirm the presence of accreting BHs in
these systems \citep[e.g.][]{Mezcua2019}. Radio telescopes have fast
survey speeds \citep[e.g.][]{Dewdney2009, Wootten2009, Perley2011,
  Haarlem2013}, the radio emission is impervious to dust obscuration
and the only source of contamination at frequencies less than 10 GHz
is star-formation activity \citep{Condon1992}, which can be mitigated
by comparing the observed radio emission to what is expected from the
measured SFR of the system
\citep[e.g.][]{Lofthouse2018,Gurkan2018,Smith2020}.  The principal bias imposed by a radio-selected sample is the sensitivity limit of the telescope in question.

In this study, we combine, for the first time, deep-wide optical and low-frequency (150 MHz) radio data, from the HSC-SSP and LOFAR \citep{vanHaarlem2013} respectively, to construct a sample of 78 dwarfs in which the radio emission far exceeds what would be expected from star formation alone, and which therefore host radio AGN. It is worth noting that the MHz frequency range is able to detect synchrotron emission with less contamination from starburst-driven thermal emission than GHz frequencies \citep[e.g.][]{Condon1992}. We use the deep optical and radio data to gain insights into the processes that likely trigger the AGN in these dwarfs, explore the properties of the AGN themselves and compare the energetics of the AGN to the expected binding energies of the gas reservoirs in dwarf galaxies to explore the plausibility of AGN feedback in this regime. 

This paper is structured as follows. In Section \ref{sec:Data}, we describe the optical (HSC-SSP) and radio (LOFAR) datasets used in this study. In Section \ref{sec:agn_selection}, we describe the construction of a sample of radio AGN in dwarf galaxies and a control sample of non-AGN matched in stellar mass and redshift. In Section \ref{sec:optical_properties}, we compare the local environments, the fraction of objects with tidal features, and the SFRs and colours in the AGN to that in their control counterparts, in order to gain insights into the properties of their host dwarf galaxies and the processes that trigger these BHs. In Section \ref{sec:agn_properties}, we combine the likely conditions in the interstellar media of dwarfs from a cosmological hydrodynamical simulation, with a semi-analytical model of the propagation of radio jets, to explore the likely jet powers, ages and accretion rates of the radio AGN in our dwarfs. We then compare the potential mechanical energy output of the AGN to the typical binding energies of the gas reservoirs to explore the plausibility of the AGN feedback in dwarf galaxies. We summarise our findings in Section \ref{sec:summary}.


\section{Data}
\label{sec:Data}


\subsection{LOFAR}

LOFAR is a wide-field radio telescope with unparalleled sensitivity
and angular resolution in the 10-240 MHz frequency range. The LOFAR
Two-metre Sky Survey \citep[LoTSS;][]{Shimwell2019} is being performed
using the High Band Antenna stations of LOFAR. LoTSS is optimised for
frequencies in the range 120–168 MHz and will eventually cover the
entire northern sky. The ELAIS-N1 field is the deepest of the
three publicly available\footnote{\url{https://lofar-surveys.org/deepfields.html}} LoTSS deep fields \citep{Sabater2021}, with a
central root mean square noise level of $\sim 20$ $\mu$Jy beam$^{-1}$
at a central frequency of 146 MHz. The resolution of the survey is 6 arcsec, with 84,862 radio sources detected at a signal-to-noise (S/N) of at least 5, within a 68 square degree area. 
The fast survey speed offered by LOFAR, due to its wide, instantaneous field of view, enables the sampling of large numbers of radio sources which, when combined with ancillary data, will enable detailed studies of star formation and AGN activity across cosmic time.


\subsection{The HSC-SSP}

The HSC-SSP is a deep-wide optical survey in four broad-band ($grizy$)
and four narrow-band filters \citep{Aihara2018a}. The Hyper
Suprime-Cam offers a 1.5 degree field of view, with a median $i$-band
seeing of $\sim 0.6$ arcsec. The survey provides Wide, Deep and
Ultra-deep layers, with areas and $r$-band point source depths of
1400, 30 and 4 deg$^2$ and $\sim 26$, 27, 28 mags respectively. The surface-brightness limit of the Deep layer (which we use
in this study) is $\sim 29$ mag arcsec$^{-2}$ \citep[e.g.][]{Huang2018}. 

The HSC-SSP provides an unprecedented combination of depth and area compared to the benchmark wide-area surveys of the past. For example, the shallowest layer of the HSC-SSP is four magnitudes deeper than standard-depth imaging from the SDSS \citep{Abazajian2009}, making this the deepest imaging in any optical survey of a comparable area, to date. The deep-wide nature of this survey makes it the best choice for obtaining, virtually for the first time, statistical samples of objects in the low-surface-brightness regime, such as dwarf galaxies at cosmological distances, and faint structures like merger-induced tidal features and intra-cluster light, out to intermediate and high-redshift \citep[e.g.][]{Kaviraj2020}. 


Here, we combine data from the LOFAR ELAIS-N1 field with the Deep layer of the HSC-SSP, to perform a statistical study of AGN in dwarfs in the redshift range $z<0.5$. We employ photometric redshifts, stellar masses and SFRs provided by the HSC-SSP \citep{Tanaka2018}, which are derived using the \textsc{DEmP} code \citep{HsiehYee2014}, applied to HSC data\footnote{The DEmP code employs polynomial fitting from 40-nearest neighbours in training sets, weighted by the separation from the galaxy in question in colour-magnitude space. The galaxies in the training sets, which include AGN, either have spectroscopic redshifts (see \citealt{HsiehYee2014} for the databases used) or highly-accurate photometric redshifts \citep[e.g.][]{Laigle2016}. The probability distributions for the derived quantities are calculated by propagating the
photometric uncertainties via Monte-Carlo simulations and the training set sampling uncertainties, using bootstrapping. In the latest HSC DEmP photo-z release, the shape/size measurements of galaxies are also included in the inputs, which improves the photometric redshift quality (Hsieh, private communication). DEmP achieves bias = -0.003, scatter = 0.019, outlier rate = 5.9 per cent for galaxies with $i <$ 24.5 (at HSC Wide depth).}. It is worth noting that, given their faintness, large spectroscopic samples of dwarfs at cosmological distances are beyond the capabilities of current instrumentation. Dwarf galaxy studies, like this one, are therefore likely to rely on photometric redshifts from deep surveys, like the HSC-SSP, for the foreseeable future.

\section{Sample selection}
\label{sec:agn_selection}

\subsection{Radio AGN in dwarf galaxies}

We first select a parent sample of low-mass objects in the HSC-SSP galaxy catalogue, which lie in the ELAIS-N1 field, have stellar masses in the range of 10$^{7.5}$ M$_{\odot}$ < M$_*$ < 10$^{9.5}$ M$_{\odot}$ and redshifts in the range $0.1 < z < 0.5$. We only consider objects that have stellar mass errors less than 0.5 dex and fractional redshift errors less than 20 per cent. The lower redshift limit is driven by the fact that there are no dwarf galaxies in the HSC-\textsc{DEmP} catalogue with redshifts less than 0.1 that also fall below our chosen fractional redshift error threshold. The upper redshift limit is driven by the fact that we wish to use visual inspection to study the presence of interaction-induced tidal features around our dwarfs (see Section \ref{sec:optical_properties}) and, given the surface-brightness limit of the HSC images, tidal features around dwarfs are unlikely to be detectable beyond $z\sim 0.5$ \citep[e.g.][]{Kaviraj2014b,Jackson2021a}.

We then cross-match this parent sample (which contains around 80,000
low-mass objects) with the LOFAR source catalogue and consider the 136
objects which have LOFAR detections with a signal-to-noise (S/N)
greater than 5 and offsets less than 1 arcsecond between the optical
and radio sources. The fraction of matches that occur simply by chance
is given by $N_d\phi\pi R^2/N_c$, where $N_d$ is the total number of
dwarfs in the overlap region between HSC-SSP and LOFAR, $\phi$ is the
average LOFAR source density (which is $9.629 \times 10^{-5}$ per
square arcsecond in the ELAIS-N1 field, see \citealt{Sabater2020}), $R$
is the maximum offset between the cross-matched optical and radio
sources (1 arcsec) and $N_c$ is the number of cross-matched sources
that have optical-radio offsets less than $R$.

\begin{figure}
    \centering
    \includegraphics[width=0.48\columnwidth]{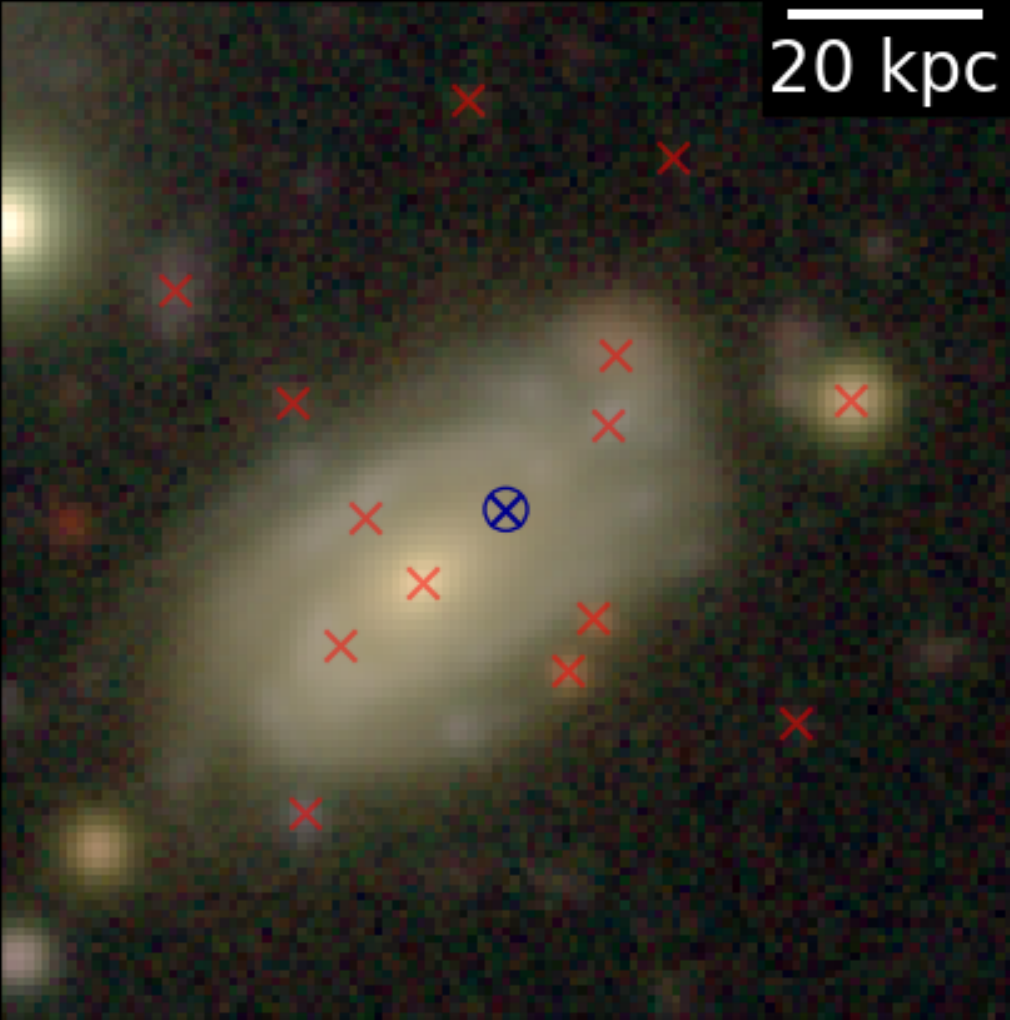}
    \includegraphics[width=0.48\columnwidth]{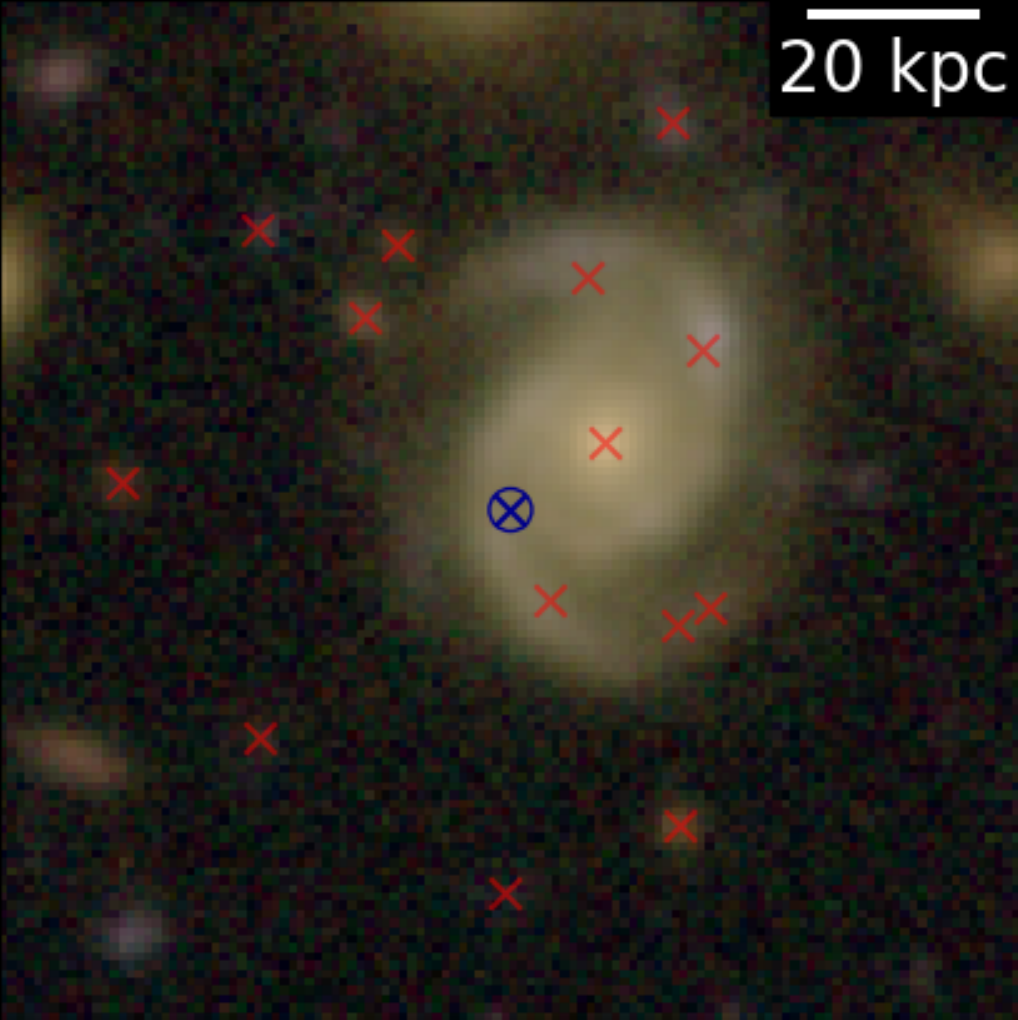}
    \caption{Example HSC colour images of two massive galaxies with bright HII regions where the optical deblender has shredded a single galaxy into multiple optical sources. As a result, the HII regions are misclassified as low mass (dwarf) galaxies. The blue cross indicates the HSC source that has been identified as a low-mass galaxy and cross-matched to a nearby LOFAR source. The red markers indicate the locations of other nearby HSC optical sources, which have also been identified as galaxies. This demonstrates how massive nearby galaxies can sometimes be shredded into several components in deep, high-resolution surveys like the HSC-SSP.}
    \label{fig:HII_region}
\end{figure}

Before calculating the chance-match fraction, we exclude a small
fraction of objects in the HSC-SSP galaxy catalogue that are actually
bright (HII) regions within massive galaxies which have been
misidentified as low-mass galaxies. Figure \ref{fig:HII_region} shows
two examples of such objects. Visual inspection of a random $\sim 5000$
low-mass objects in the HSC-SSP Deep survey, in our mass range of
interest, indicates that the fraction of such objects that are
actually HII regions and not dwarf galaxies is $\sim 2$ per cent.
Taking these misidentifications into account, the chance match
fraction is $\sim 17$ per cent. 

\begin{figure}
    \centering
    \includegraphics[width=0.32\columnwidth]{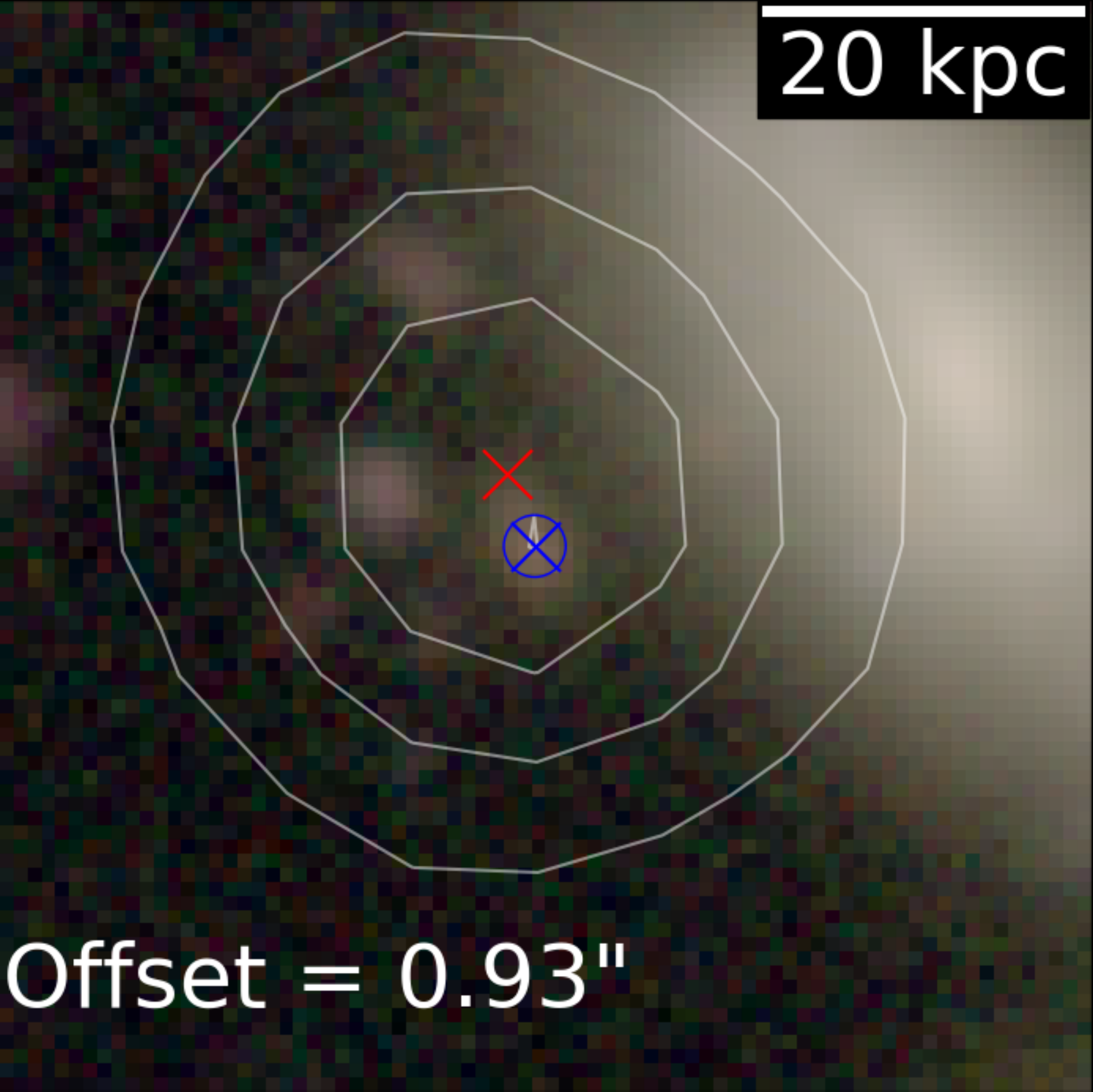}
    \includegraphics[width=0.32\columnwidth]{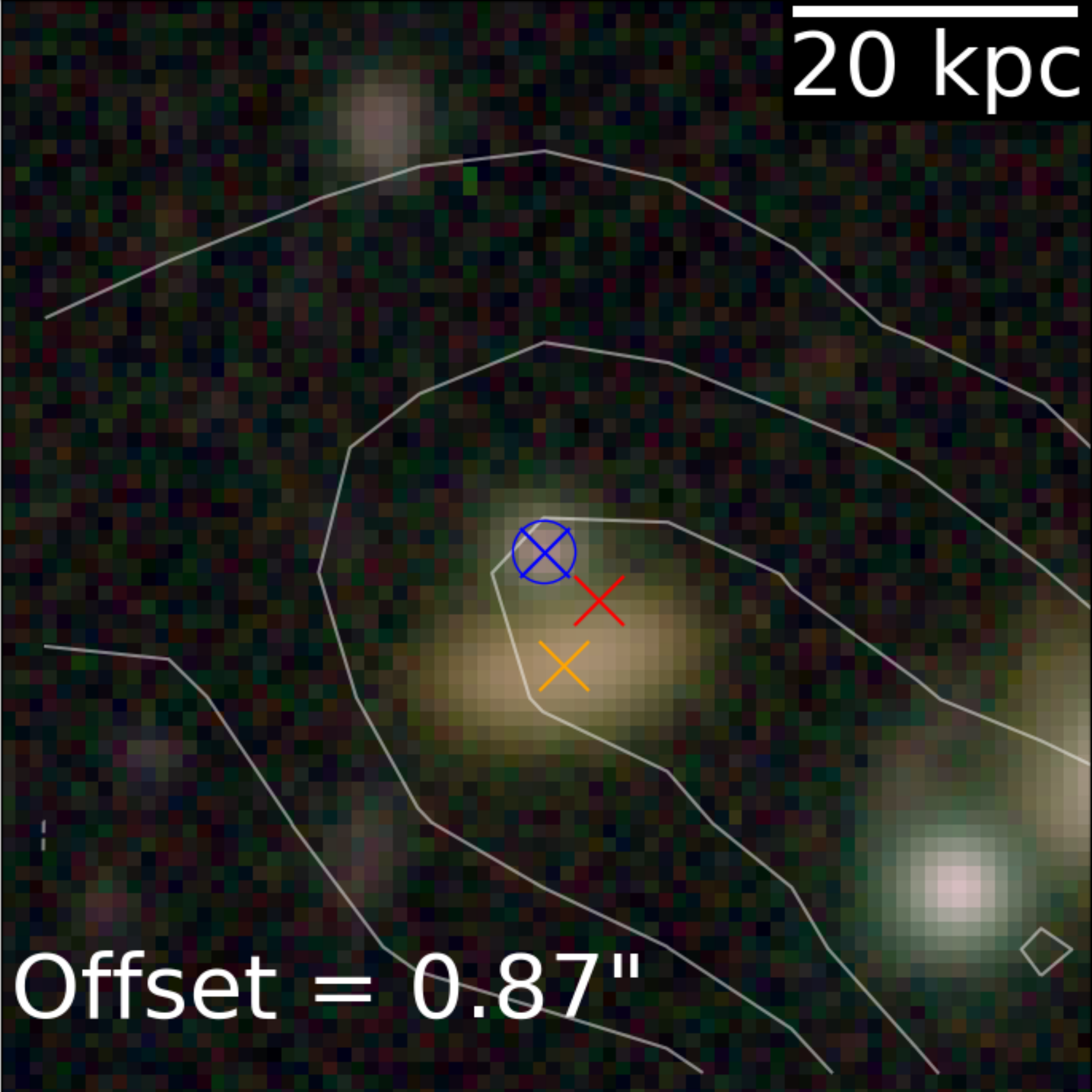}
    \includegraphics[width=0.32\columnwidth]{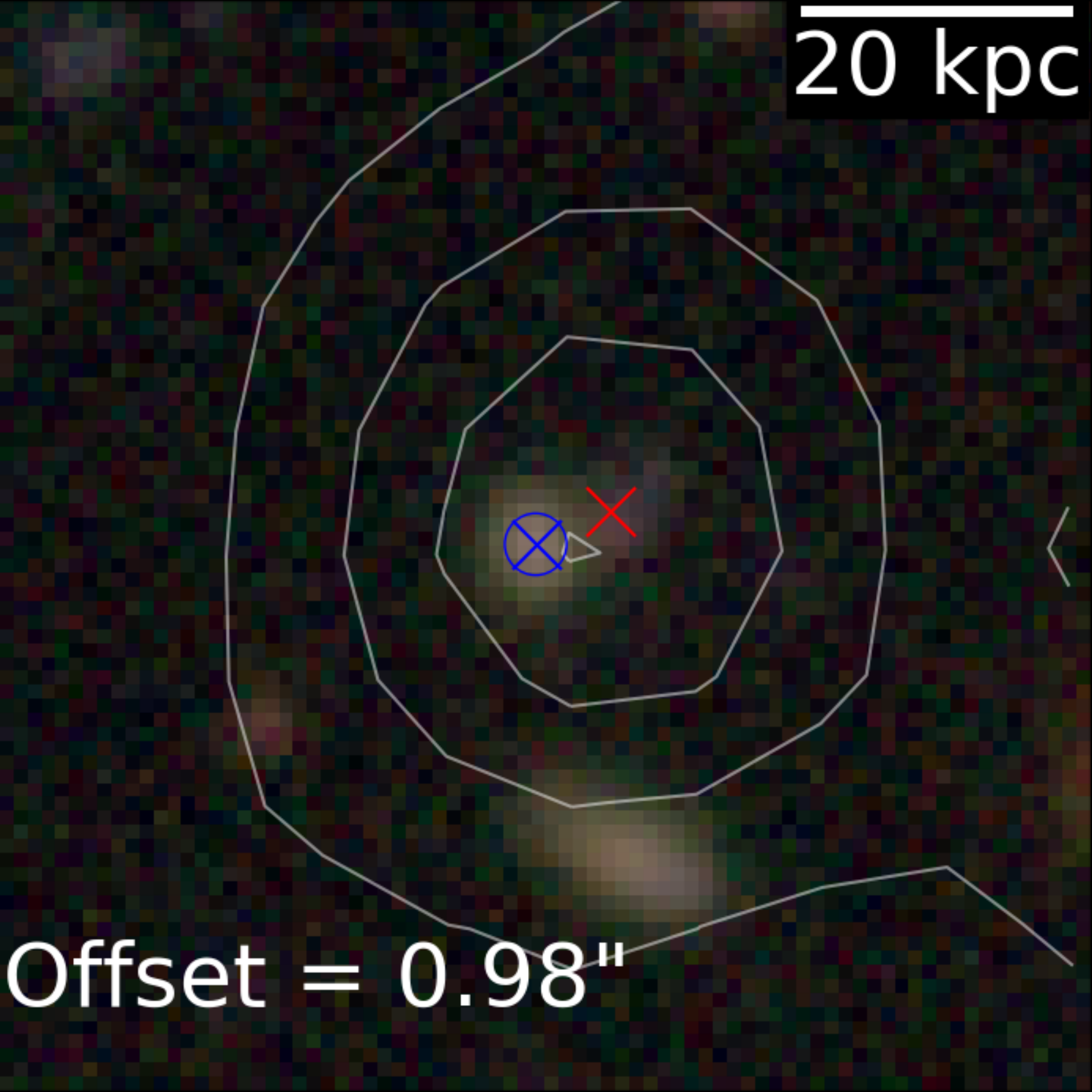}
    \caption{HSC colour images of three dwarf galaxies, with LOFAR matches within 1 arcsecond, where the radio emission could either be contaminated by nearby neighbours or be due to background sources. These objects are removed from our final sample via visual inspection of such images. The red and blue crosses indicate the locations of the radio and optical source centroids respectively. The extent of the radio emission is demonstrated using the white contours. The contour levels in each of the above images (left to right) increase linearly by 200, 33 and 100 $\mu$Jy beam$^{-1}$, with a maximum of 750, 130, and 380 $\mu$Jy beam$^{-1}$ respectively.  \textbf{Left:} Given the large optical-radio offset, which places the radio centroid outside the main body of the dwarf, a background source could be responsible for the radio emission in this object. \textbf{Middle:} The close proximity of a nearby galaxy (orange), along with the extended radio emission to the south east suggests that the radio emission coincident with the dwarf could be contaminated by these objects. \textbf{Right:} The radio centroid is situated outside the body of the dwarf, and appears coincident with a faint nearby optical source to the east, which could potentially be the dominant source of the radio emission.
    }
    \label{fig:contaminants}
\end{figure}

\begin{figure}
    \centering
    \includegraphics[width=0.45\columnwidth]{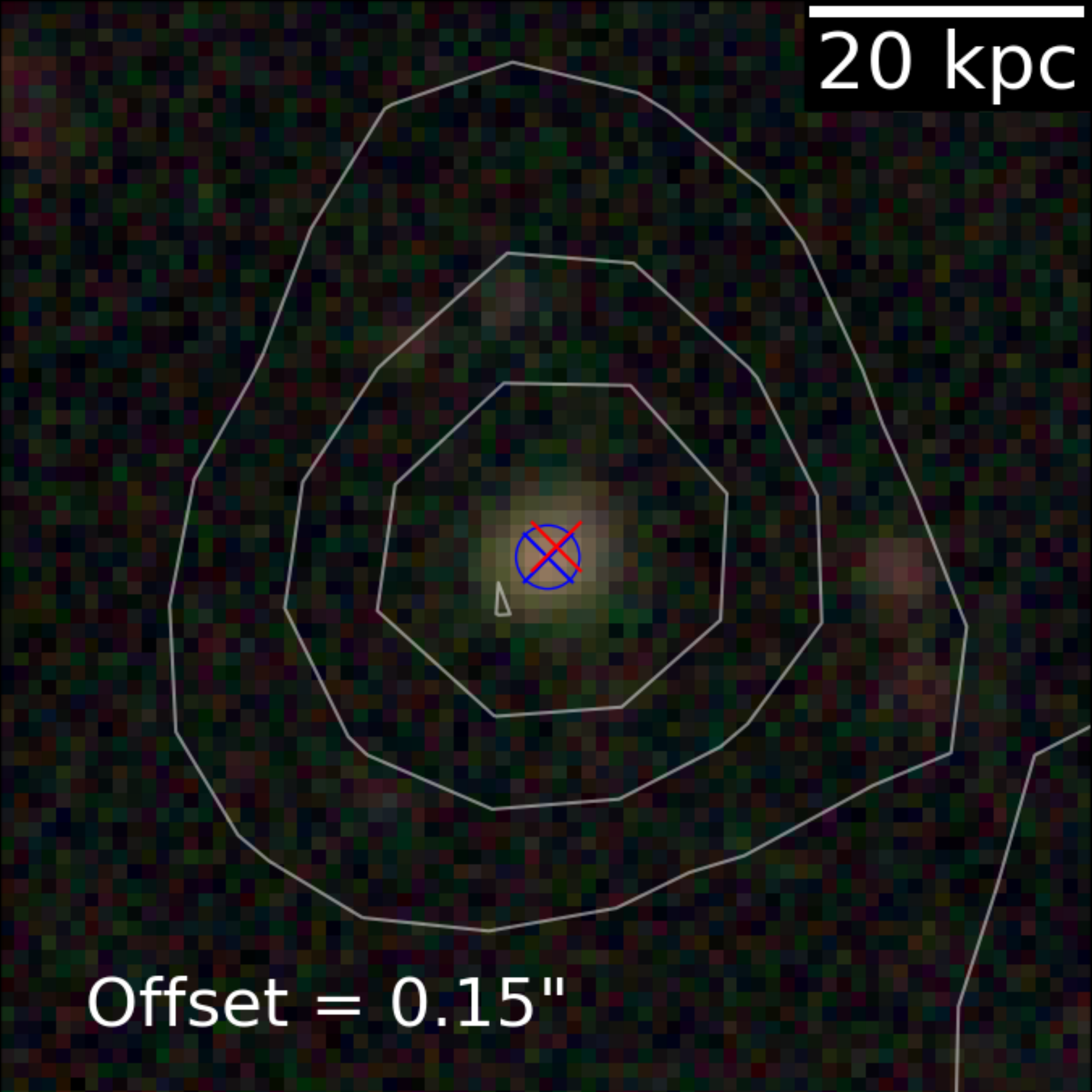}
    \includegraphics[width=0.45\columnwidth]{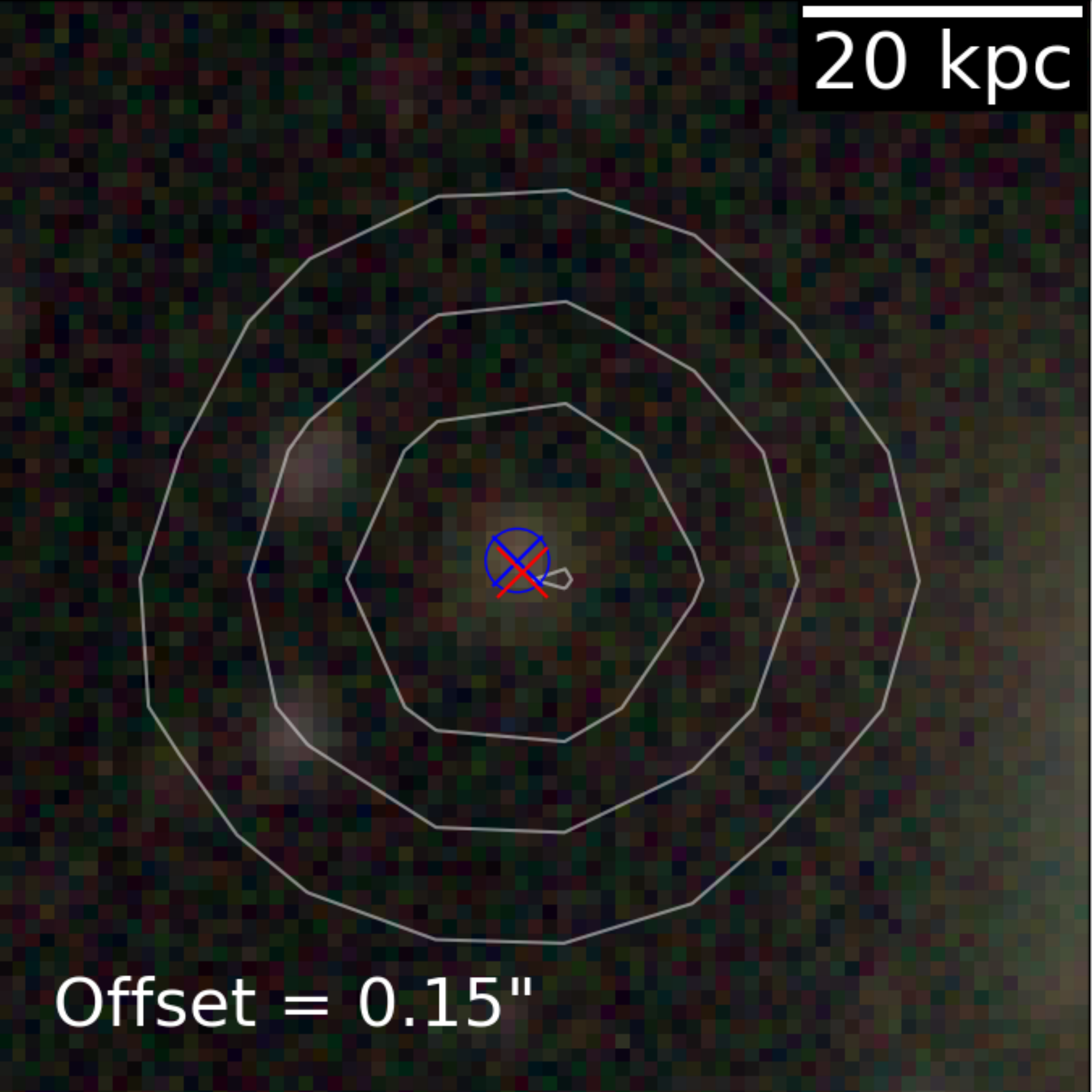}
    \includegraphics[width=0.45\columnwidth]{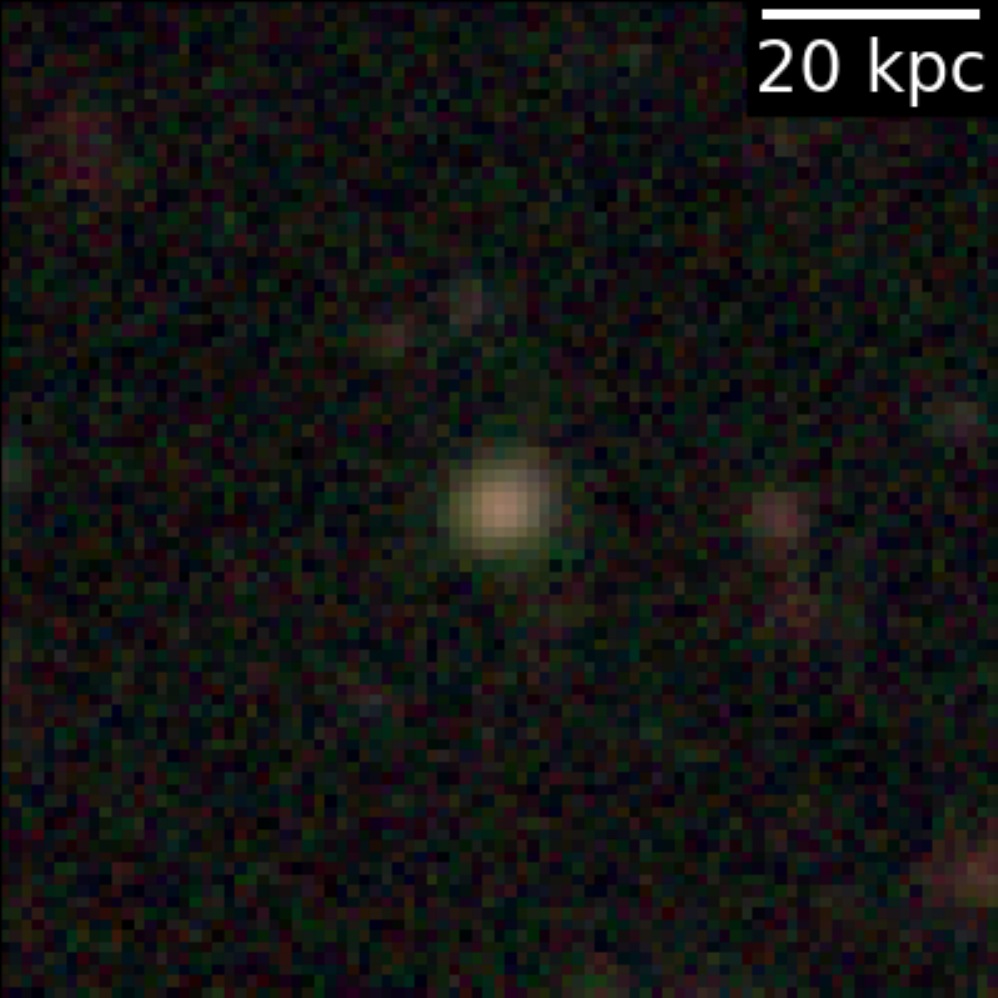}
    \includegraphics[width=0.45\columnwidth]{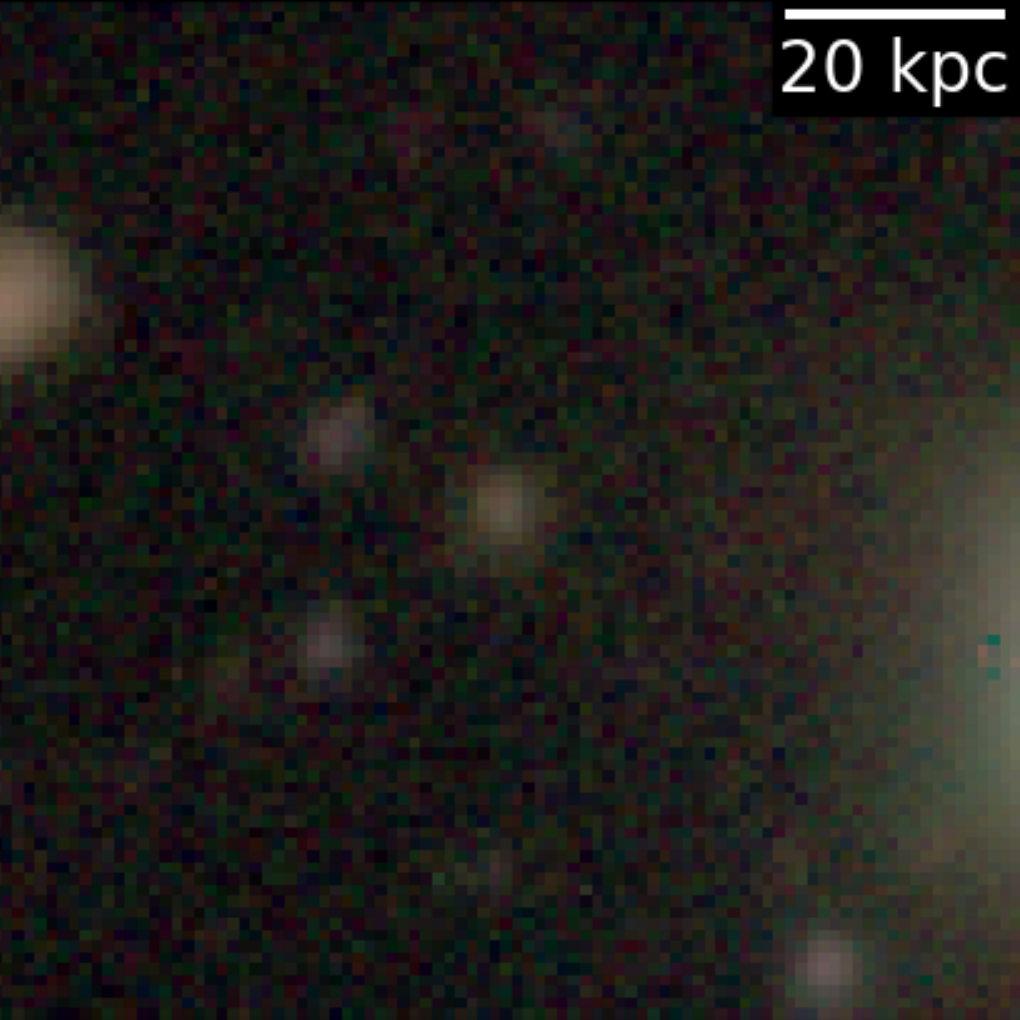}

    \caption{\textbf{Top row:} HSC colour images of two typical dwarf galaxies in our final AGN sample, which are confirmed, via visual inspection, to be cross-matched correctly with their radio source counterparts. The red and blue crosses indicate the locations of the radio and optical source centroids respectively. The extent of the radio emission is demonstrated using the white contours. The contour levels in each of the images above (left to right) increase linearly by 70 and 200 $\mu$Jy beam$^{-1}$, with a maximum of 250 and 820 $\mu$Jy beam$^{-1}$ respectively. In all cases the radio centroid sits within the body of the dwarf and there are no potential contaminants within close proximity to the object. \textbf{Bottom row:} HSC colour images of two dwarf galaxies in the control samples selected for the AGN in the top row. As noted in Section \ref{sec:controls}, control-sample dwarfs are stellar mass and redshift-matched to their corresponding AGN.}
    \label{fig:optical_radio_images}
\end{figure}

To obtain as clean a sample as possible, we visually inspect all
HSC-SSP objects in our mass range of interest that have a radio match
within 1 arcsecond, using HSC colour images overlaid with LOFAR radio
contours. We use this process to remove objects with potential nearby
radio contaminants and those that are possible background sources (see
Figure \ref{fig:contaminants}). Figure  \ref{fig:optical_radio_images}
shows examples of the LOFAR-detected dwarfs that comprise our final AGN sample (top row) and also dwarfs from the control sample (bottom row).




\begin{figure}
    \centering
    \includegraphics[width=\columnwidth]{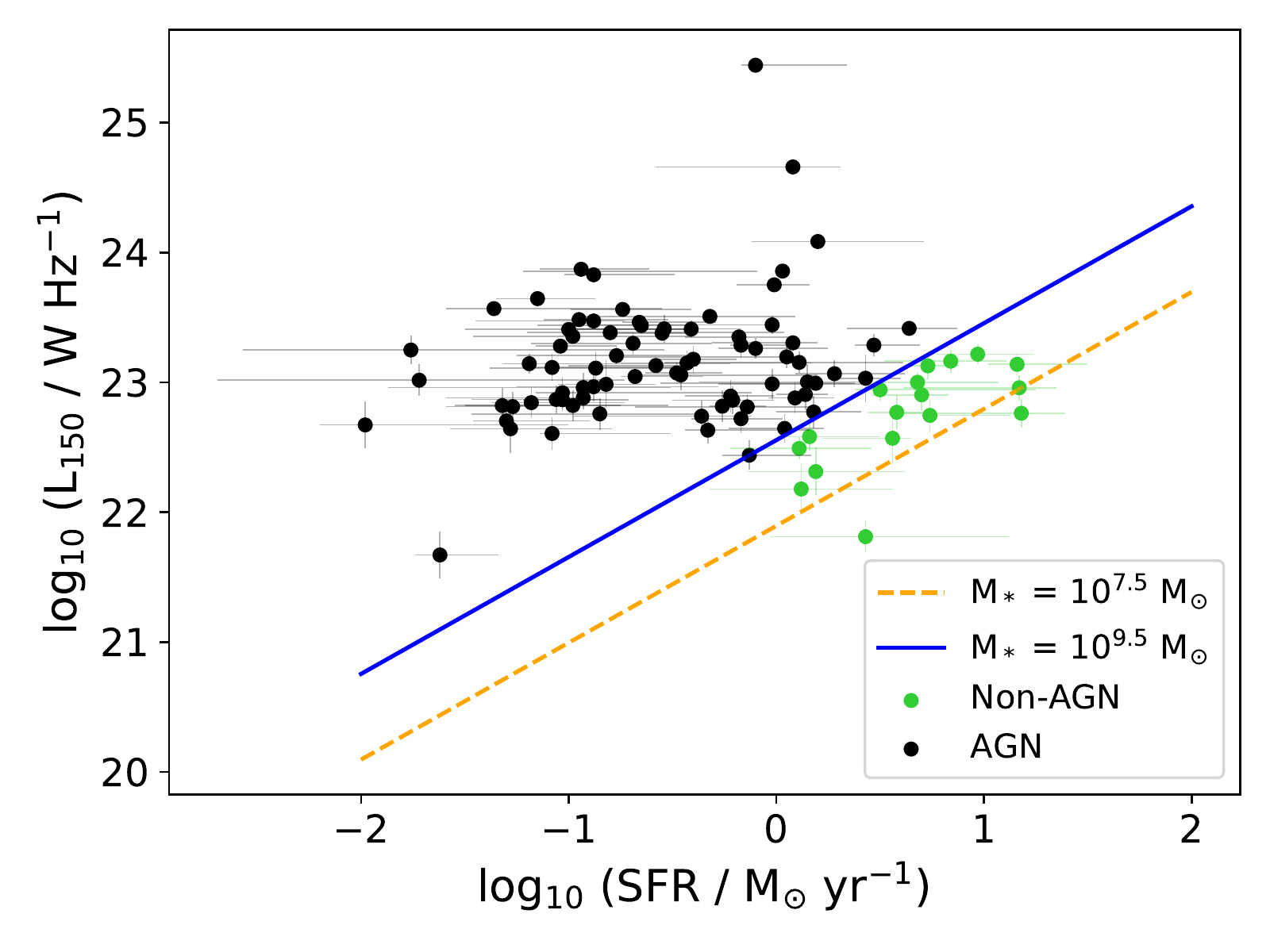}
     \includegraphics[width=\columnwidth]{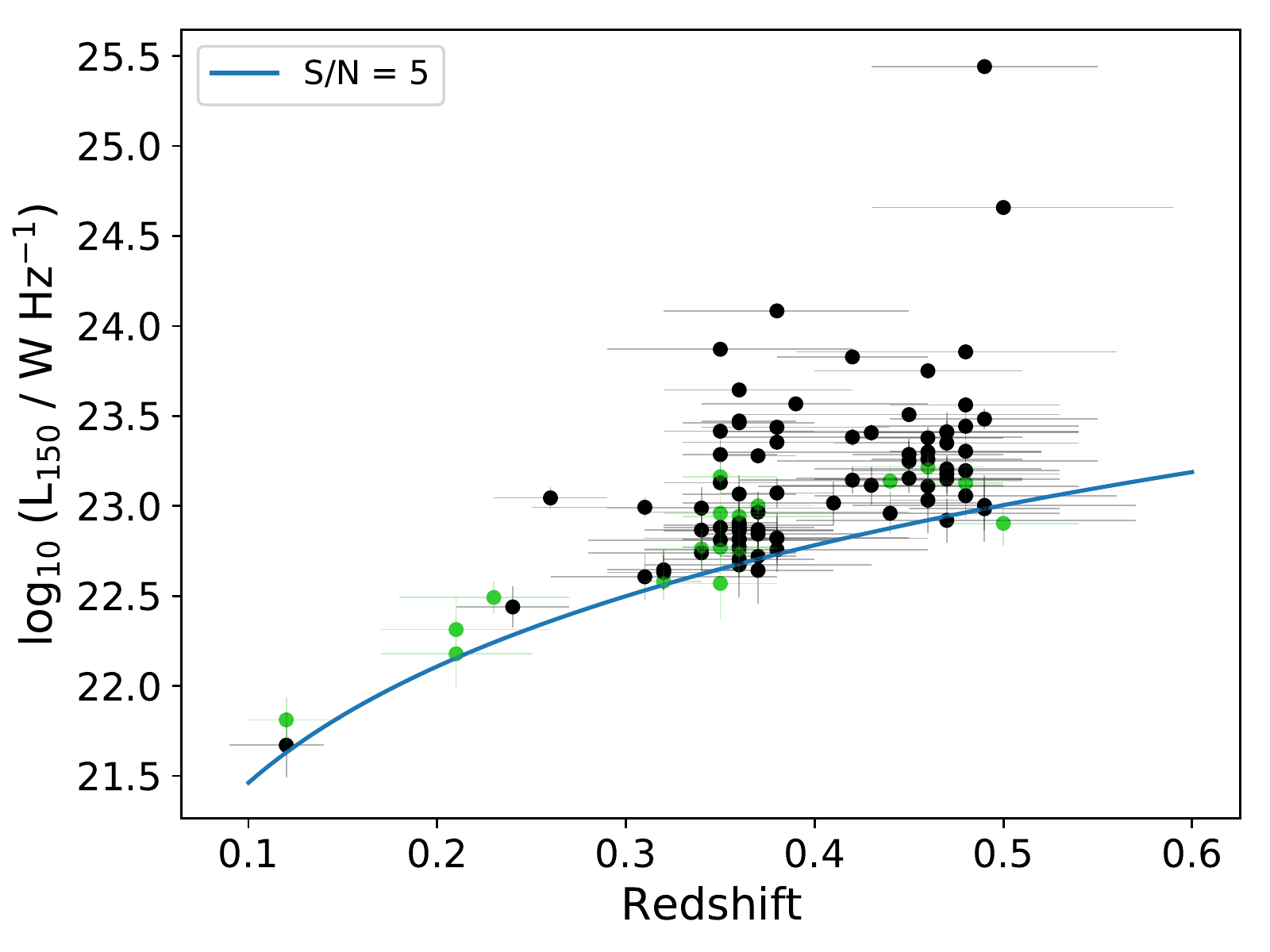}
    \includegraphics[width=\columnwidth]{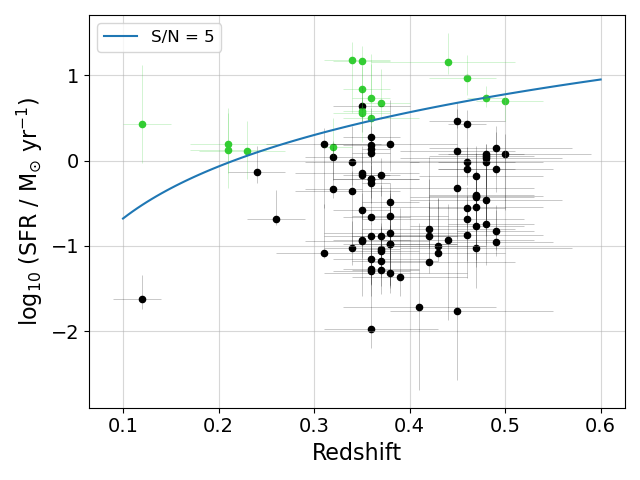}
    \caption{\textbf{Top:} L$_{150}$ vs SFR for our final sample of dwarf galaxies which have LOFAR counterparts. The solid blue and dashed orange lines show L$_{150}$ - SFR relations from \citet{Smith2020} for two stellar masses (orange = $10^{7.5}$ M$_\odot$, blue = $10^{9.5}$ M$_\odot$) with a +0.5 dex excess added to the median relation (which is our criterion for selecting AGN). Black points represent AGN (i.e. objects where the radio luminosity is more than 0.5 dex greater than the L$_{150}$ that corresponds to the SFR), while green points indicate dwarfs where the L$_{150}$ is consistent with star formation (i.e. within 0.5 dex of the L$_{150}$ that corresponds to the SFR). \textbf{Middle and bottom:} The L$_{150}$ (middle) and SFR (bottom) of our LOFAR-detected dwarfs as a function of redshift. Black and green points indicate AGN and non-AGN respectively. The light blue curves indicate the 5 sigma detection threshold of LOFAR. 
    }
    \label{fig:relation}
\end{figure}

\begin{figure}
    \centering
    \includegraphics[width=0.99\columnwidth]{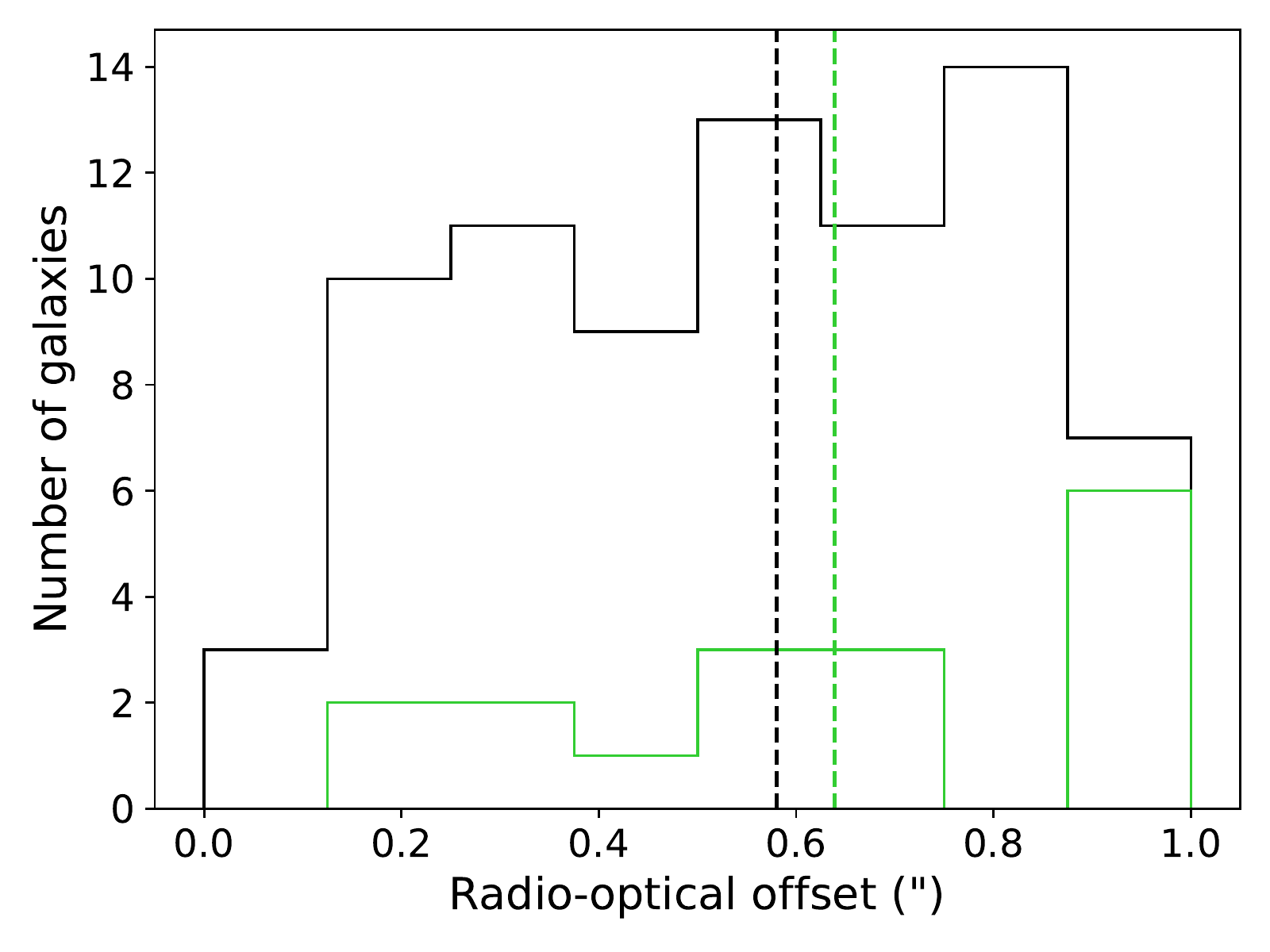}
    \includegraphics[width=0.99\columnwidth]{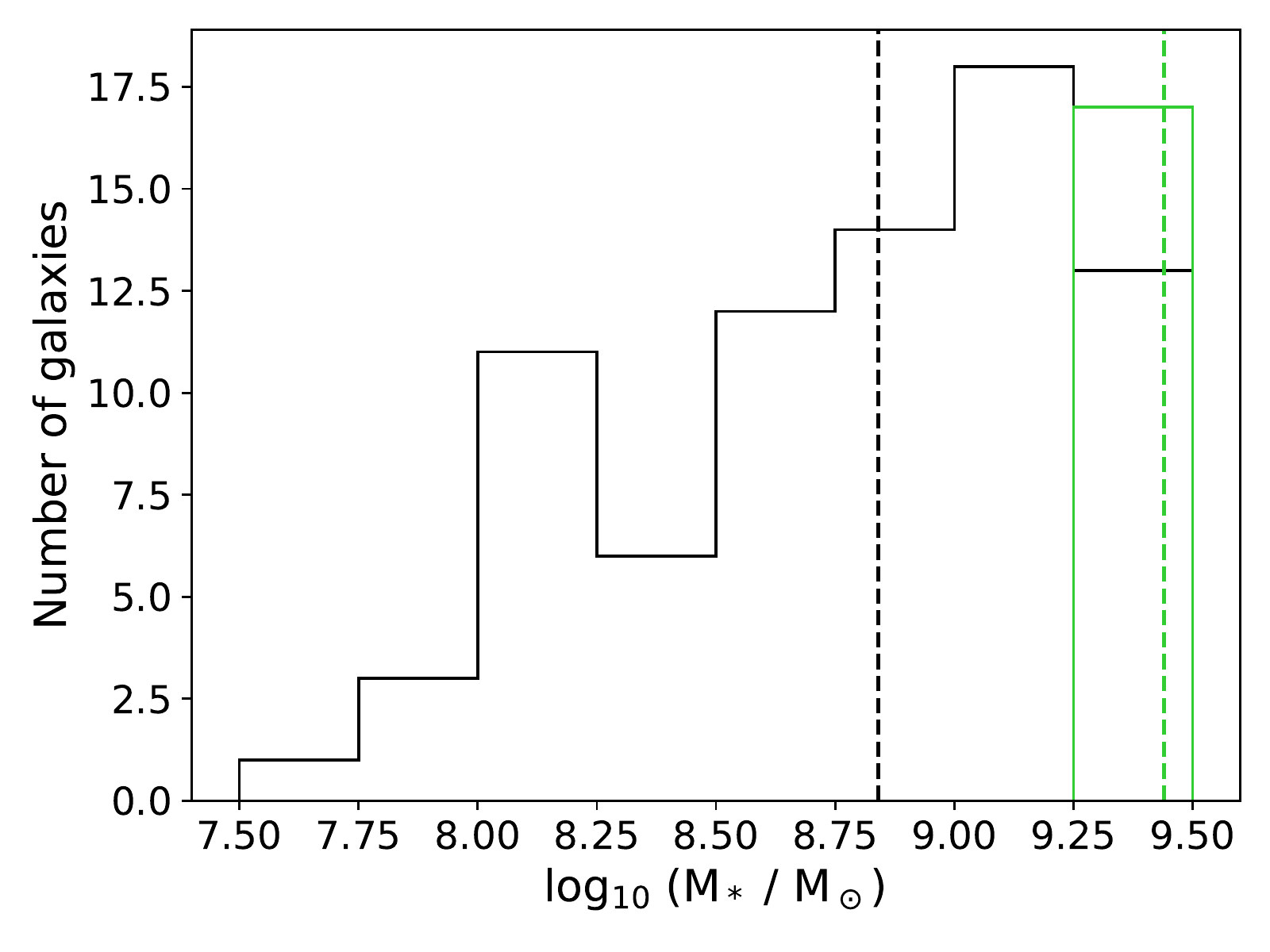}
    \includegraphics[width=0.99\columnwidth]{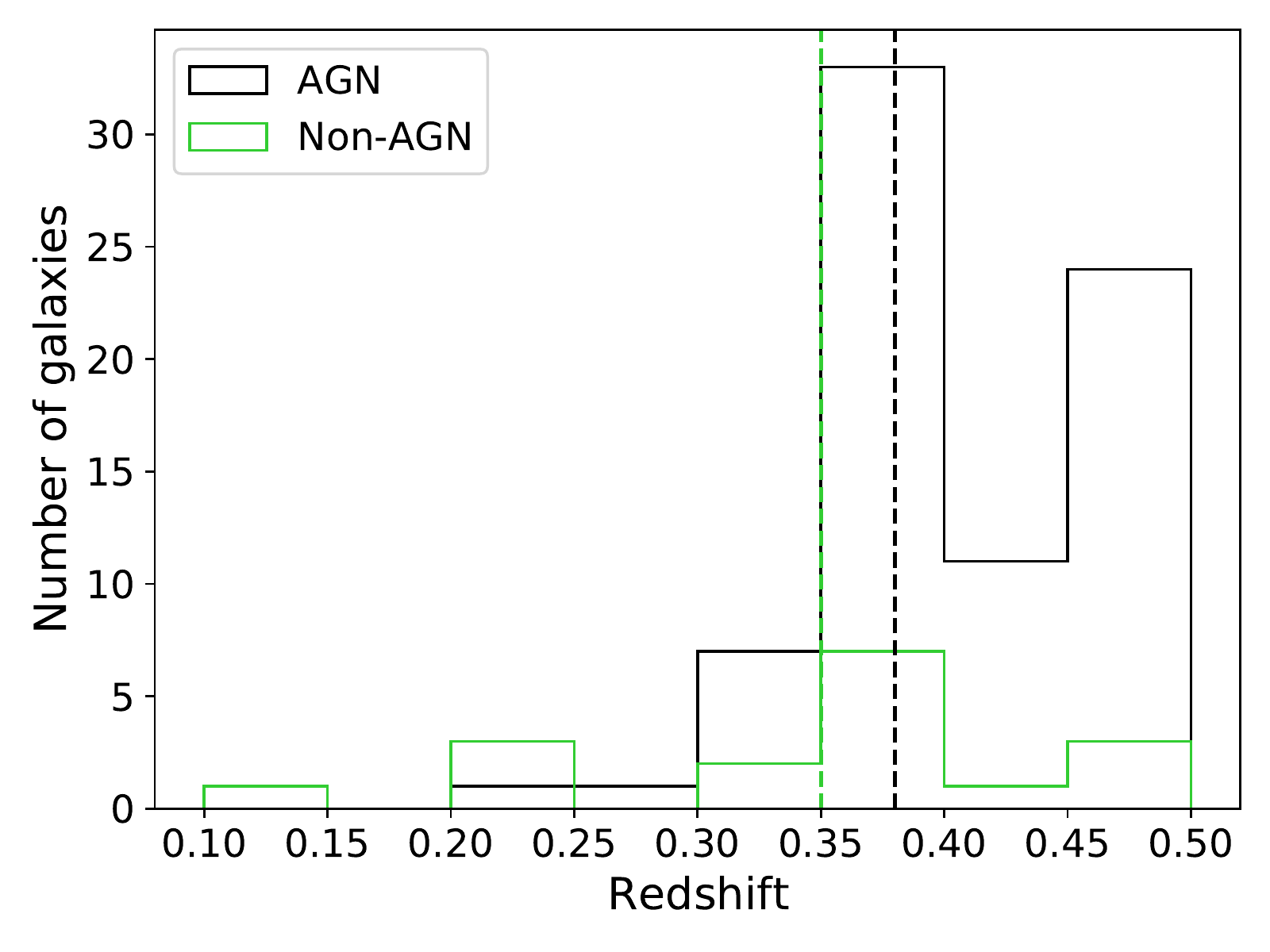}
    \caption{Distributions of radio-optical offsets (top), stellar
      masses (middle) and redshifts (bottom) for the AGN (black) and
      non-AGN (green). Median values are shown using the dashed
      vertical lines.}    \label{fig:sampledistributions}
\end{figure}

Finally, we identify AGN in our dwarf galaxy sample through their excess radio emission, beyond what is expected due to star formation alone, using the L$_{150}$ - SFR relations from \citet{Smith2020} for low-mass galaxies (where L$_{150}$ is the LOFAR 150 MHz luminosity). We first use the HSC-SSP SFR to calculate the L$_{150}$ emission that is contributed by star formation using the \citet{Smith2020} L$_{150}$ - SFR relation. We then identify systems that host AGN as those where the L$_{150}$ measured by LOFAR exceeds the L$_{150}$ predicted from the star formation rate by 0.5 dex or more. This indicates that the radio emission is more than three times that which can be accounted for by star formation activity alone, indicating the presence of another radio source, which can only plausibly be an AGN. We note that there is no trend between the stellar mass of the dwarfs and the L$_{150}$ excess.
As shown in the top panel of Figure \ref{fig:relation}, this results in 78 dwarf galaxies that host AGN (black points) and 17 that are detected by LOFAR but where the radio luminosity is consistent with star formation (green points). In the middle and bottom panels of this figure, we plot the L$_{150}$ and SFR of our LOFAR-detected dwarfs as a function of redshift, and indicate the LOFAR 5 sigma detection thresholds in these quantities. 

The distributions of radio-optical offsets, stellar masses and redshifts for our final sample of dwarfs are shown in Figure \ref{fig:sampledistributions}. While there are some dwarfs which do not host AGN but are detected by LOFAR, these are all clustered within 0.25 dex of the upper limit of our mass range (middle panel of Figure \ref{fig:sampledistributions}), with a median stellar mass that is 0.6 dex larger than that of the AGN. Thus, while the majority (82 per cent) of LOFAR-detected dwarfs which have M$_*$ < 10$^{9.5}$ M$_{\odot}$, at the redshifts being considered, are AGN, the AGN fraction of LOFAR-detected dwarfs is essentially 100 per cent at M$_*$ < 10$^{9.25}$ M$_{\odot}$ at these redshifts. In other words, given the SFRs in dwarf galaxies at such epochs, these systems typically need a radio AGN to actually be detectable by LOFAR.



\subsection{Control sample}
\label{sec:controls}

We construct a control sample of non-AGN dwarf galaxies which is
matched to our AGN population in stellar mass and redshift. We first
exclude all HSC-SSP galaxies with a LOFAR radio source detection
within 6 arcsec. For each of the 78 dwarf AGN galaxies, we then
randomly select three dwarfs, each of which has a redshift within 0.01
and a stellar mass within 0.05 dex of the AGN in question and identical error constraints in mass and redshift as those used to select the LOFAR-selected dwarfs. We visually inspect this control sample in order to remove objects which are not bona-fide dwarfs (e.g. HII regions). If such objects are found they are replaced by another randomly drawn dwarf, with the same selection criteria as those described above.
A two sample Kolmogorov–Smirnov (KS) test for the redshift and stellar
mass distributions of the AGN and control samples both result in a
$p$-value close to 1.

\section{Local environments, fractions of interacting galaxies and star formation properties}
\label{sec:optical_properties}

We begin our analysis by studying the environments of our AGN. In the absence of spectroscopic redshifts, we consider projected densities, using 2D nearest-neighbour distances to massive galaxies (M$_*$ > 10$^{10.5}$ M$_{\odot}$) which have redshifts within the one-sigma limits of the redshift of each dwarf galaxy. Projected densities, quantified using such nearest-neighbour distances, correlate with the true 3D density, and the relative behaviours in projection are preserved in three dimensional space \citep[e.g.][]{Shattow2013}. We calculate 2D distances to the first, second, third, fifth and tenth nearest massive galaxies. These distances trace both the very local neighbourhood and the more large-scale environment around each dwarf. We then compare the distributions of these distances for the AGN and controls, to probe whether differences exists in the environments of the AGN and the control galaxies. 

\begin{figure}
    \centering
    \includegraphics[width=\columnwidth]{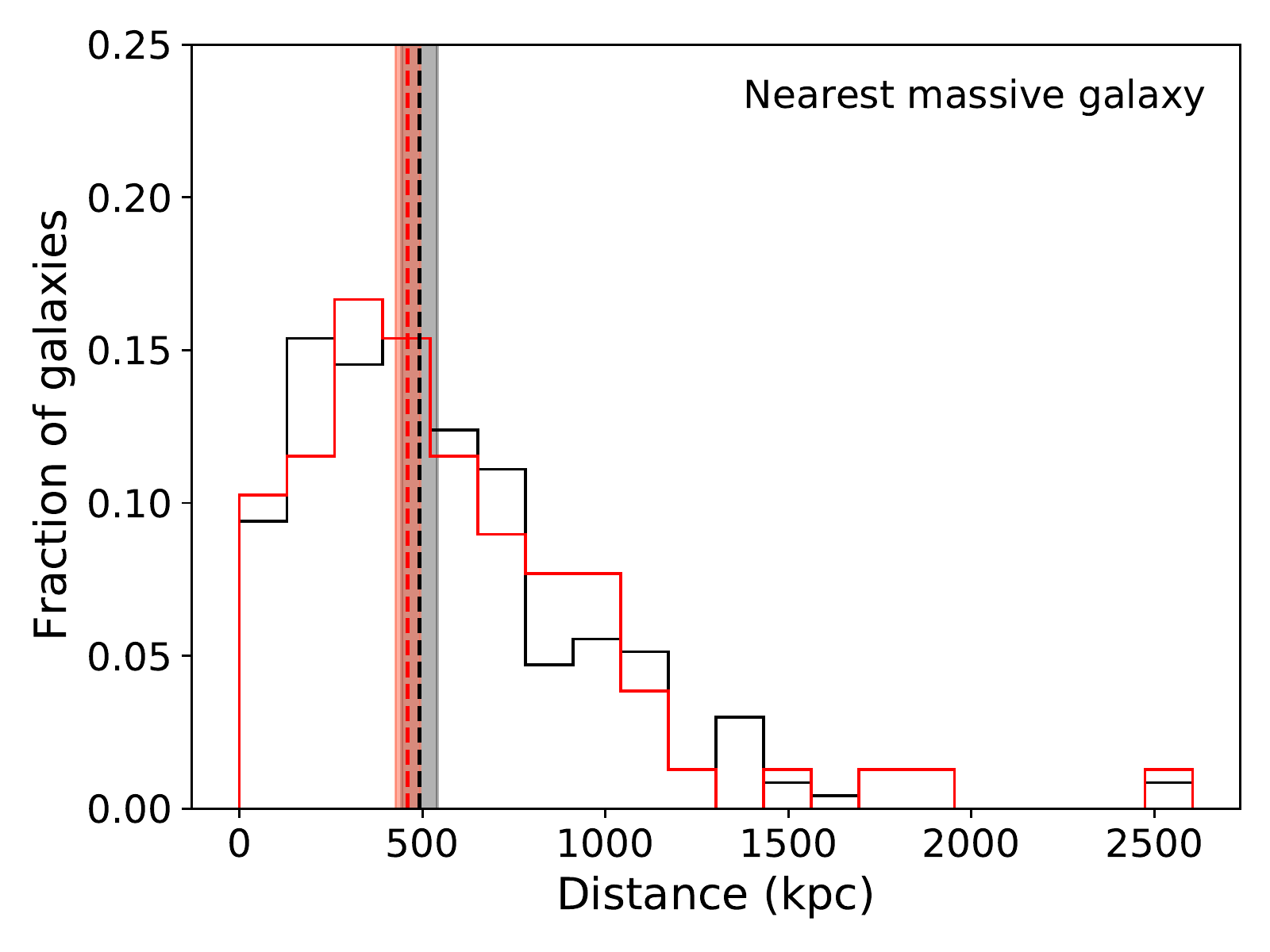}
    \includegraphics[width=\columnwidth]{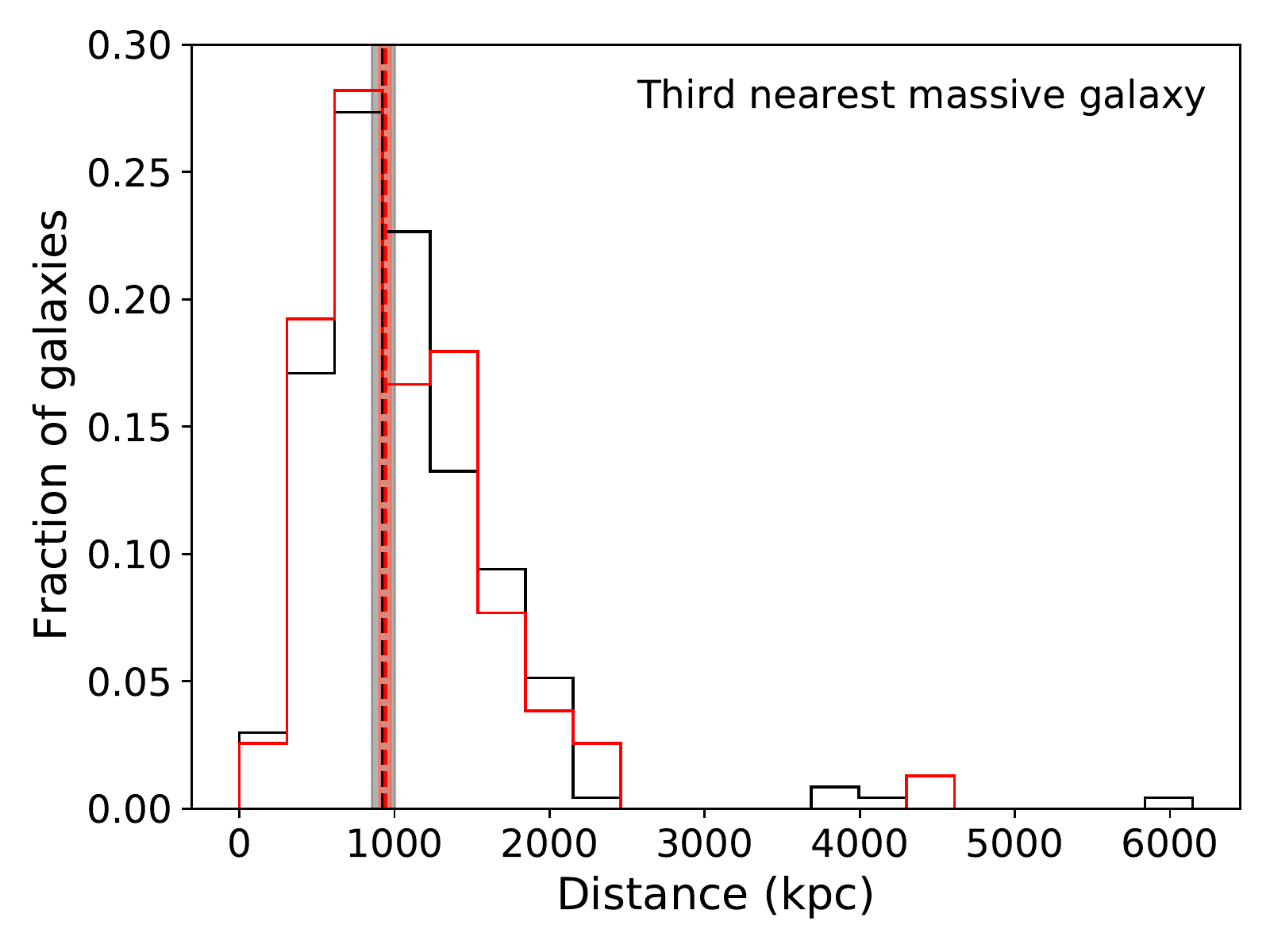}
    \includegraphics[width=\columnwidth]{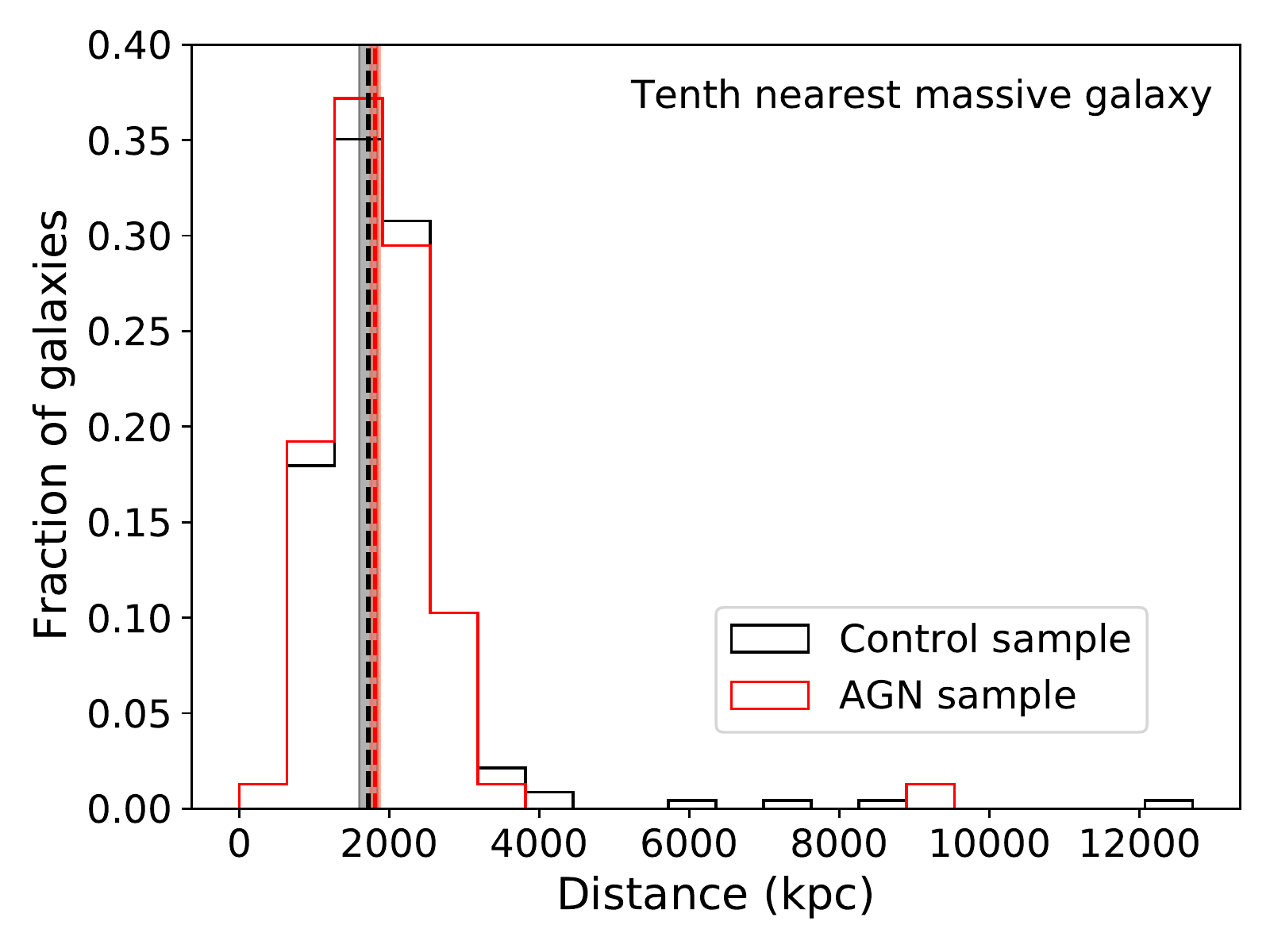}
    \caption{2D distances (calculated using RA and DEC only within a redshift slice) to the
      nearest (top panel), third nearest (middle panel) and tenth
      nearest (bottom panel) massive galaxies (M$_*$ > 10$^{10.5}$
      M$_{\odot}$) for dwarfs in the AGN (black) and control (red)
      samples. We only consider massive galaxies which have redshifts
      within the one-sigma limits of the redshift of the dwarf galaxy
      in question. Medians and the corresponding bootstrapped
      uncertainties are shown using the vertical dashed lines and shaded
      regions respectively. The median values for the AGN and controls
      are consistent within the uncertainties for each of the
      distances considered. Both the very local neighbourhoods and the large-scale environments of the AGN and controls appear similar, and the triggering of these AGN is unlikely to be driven by environmental factors.
    }
    \label{fig:nearestmg}
\end{figure}

Figure \ref{fig:nearestmg} indicates that the 2D distances to the nearest, third nearest and tenth nearest massive galaxies do not show any differences between the AGN and the control samples. Median values are indicated using the dashed vertical lines and bootstrapped errors on the medians are indicated using the shaded regions. The median values for the AGN and control populations are consistent within the errors. Furthermore, a Mann-Whitney U test \citep{mannwhitney} yields $p$-values greater than 0.05 in all cases, indicating that there are no statistically significant differences in the distributions of distances for the AGN and controls. We note that the second and fifth nearest distances also show the same behaviour (although we omit these from Figure \ref{fig:nearestmg} for clarity). Both the very local neighbourhoods and the large-scale environments of the AGN and the control population therefore appear similar. Thus, the triggering of these AGN is unlikely to be driven by environmental factors, e.g. higher-density regions that can cause elevated rates of gas accretion in the AGN \citep[e.g.][]{Jackson2021a} and/or induce stronger tidal fields which can trigger the internal gas reservoirs of these systems \citep[e.g.][]{Martin2019,Jackson2021b}.

\begin{figure}
    \centering
    \includegraphics[width=0.48\columnwidth]{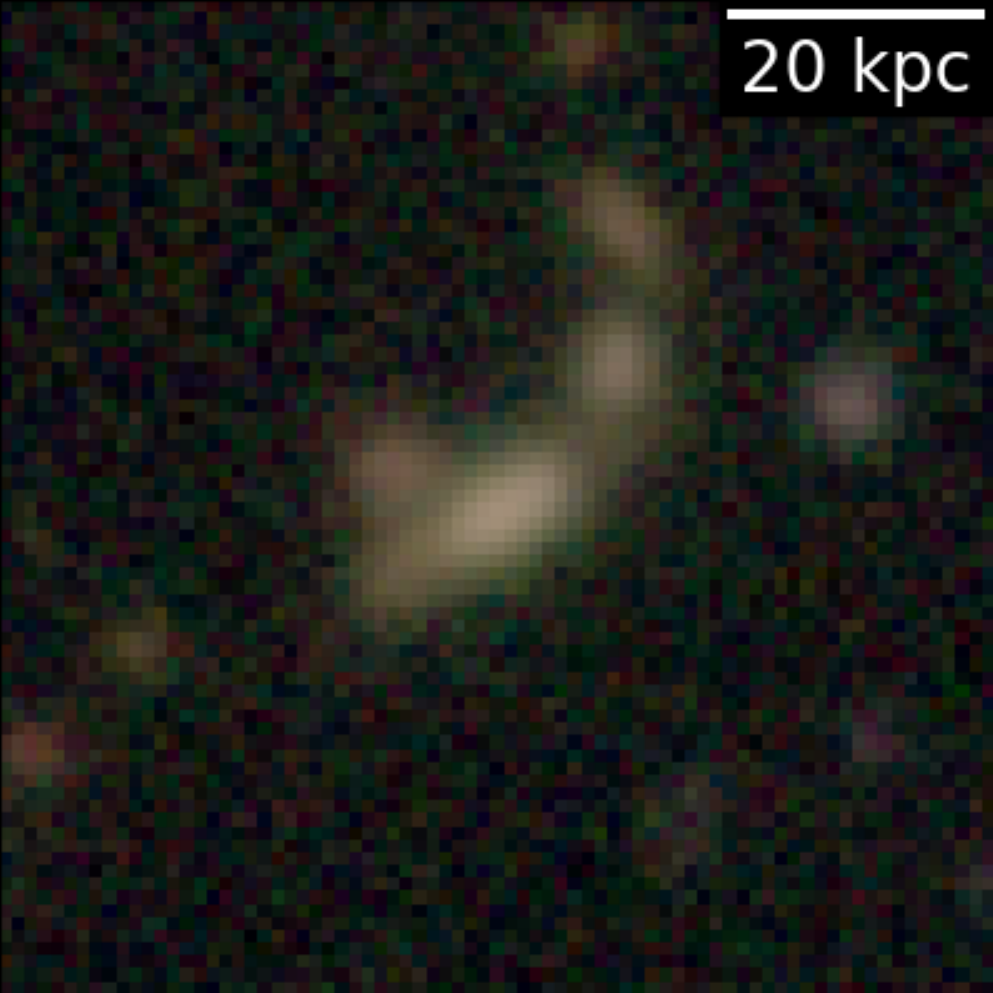}
    \includegraphics[width=0.48\columnwidth]{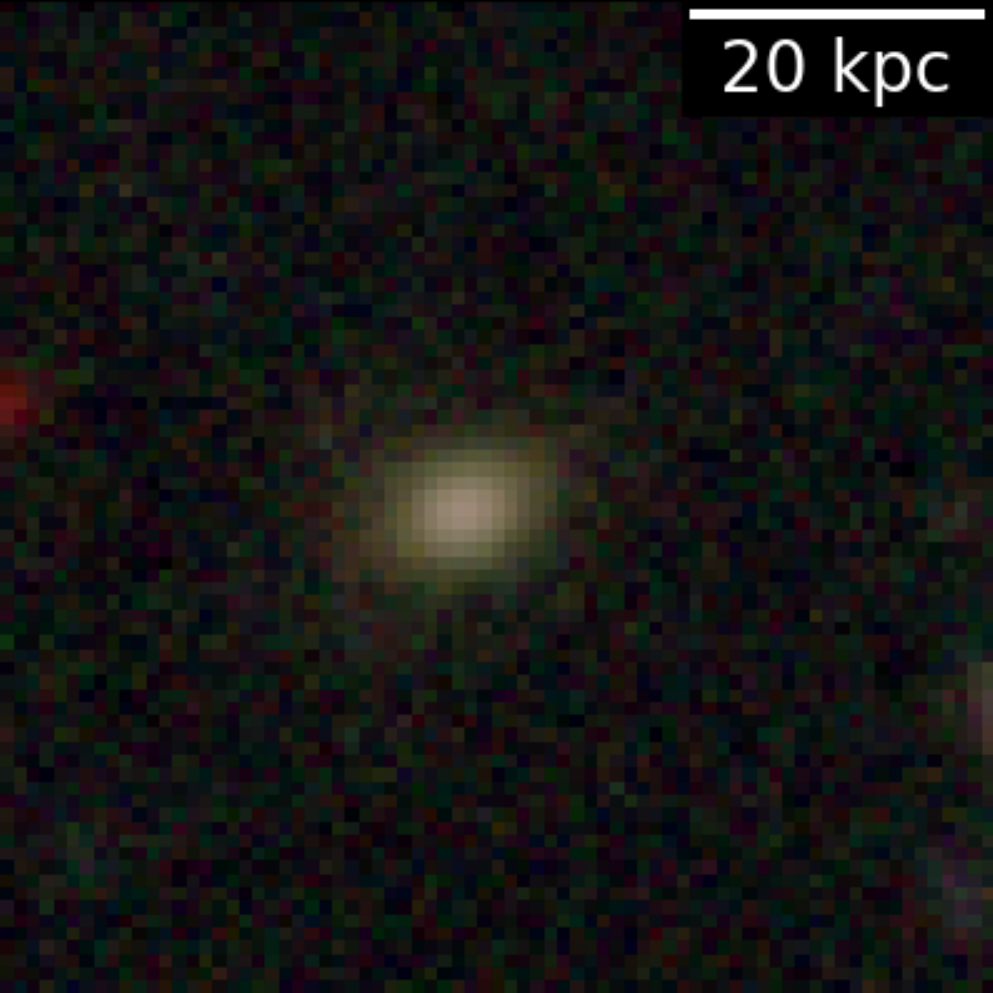}
    \caption{HSC colour images of a dwarf galaxy with (left) and without (right) tidal features that are indicative of an interaction with other galaxies.}
    \label{fig:tidalfeatures}
\end{figure}

The fraction of objects that show tidal features in the HSC images
(Figure \ref{fig:tidalfeatures}) is 6.4$_{-1.8}^{+3.9}$ per cent in
the AGN compared to 4.7$_{-1.0}^{+1.8}$ per cent in their control
counterparts\footnote{It is worth noting that these fractions agree
  well with the fraction of dwarfs that are predicted to be morphologically disturbed in
  cosmological simulations at late epochs \citep[$\sim 5$ per cent,][]{Martin2021}.}, where the uncertainties are calculated following \citet{Cameron2011}\footnote{\citet{Cameron2011} provides a method of calculating accurate Bayesian binomial confidence intervals using the quantiles of the beta distribution. These are typically more accurate than simpler methods, like using the normal approximation, which can misrepresent the degree of statistical uncertainty that is present under the sampling conditions often encountered in astronomical surveys.}. The low tidal fraction, coupled with the fact that the tidal fractions are consistent, within uncertainties, in the AGN and control samples, suggests that the AGN are not being triggered by mergers with lower mass galaxies or fly-bys with massive companions \citep[which can also produce tidal features in dwarfs, see][] {Jackson2021b}. The latter point is consistent with the fact that, while fly-by induced tidal features in dwarfs only appear when they pass very close (at distances less than 100 kpc, see e.g. \citealt{Jackson2021b}) to massive galaxies, the 2D distances of the dwarfs to their nearest massive galaxies are all much larger (the fraction of nearest massive galaxies that are within 100 Mpc of a dwarf is $\sim 6$ percent for both the AGN and the controls). 

As we show in Section \ref{sec:agn_properties} below, our AGN are
likely to be driven by activity that has been triggered recently, because the radio luminosities in older AGN would fall below the detection limit of LOFAR at the redshifts probed in this study. If these young AGN were merger-driven, then the triggering mergers would be recent and the tidal features would have had very little time to fade, making them more easily detectable and making it more likely that the tidal fraction would be elevated in the AGN compared to the controls. However, since the tidal fractions are consistent between the two samples, this reinforces the argument that our AGN are not driven by interactions. It is worth noting that the low tidal fraction, and the apparent irrelevance of interactions in triggering AGN found here, is in agreement with the conclusions of \citet{Kaviraj2019}, who studied AGN in a large sample of dwarf galaxies using infrared data.  

\begin{figure}
    \centering
    \includegraphics[width=\columnwidth]{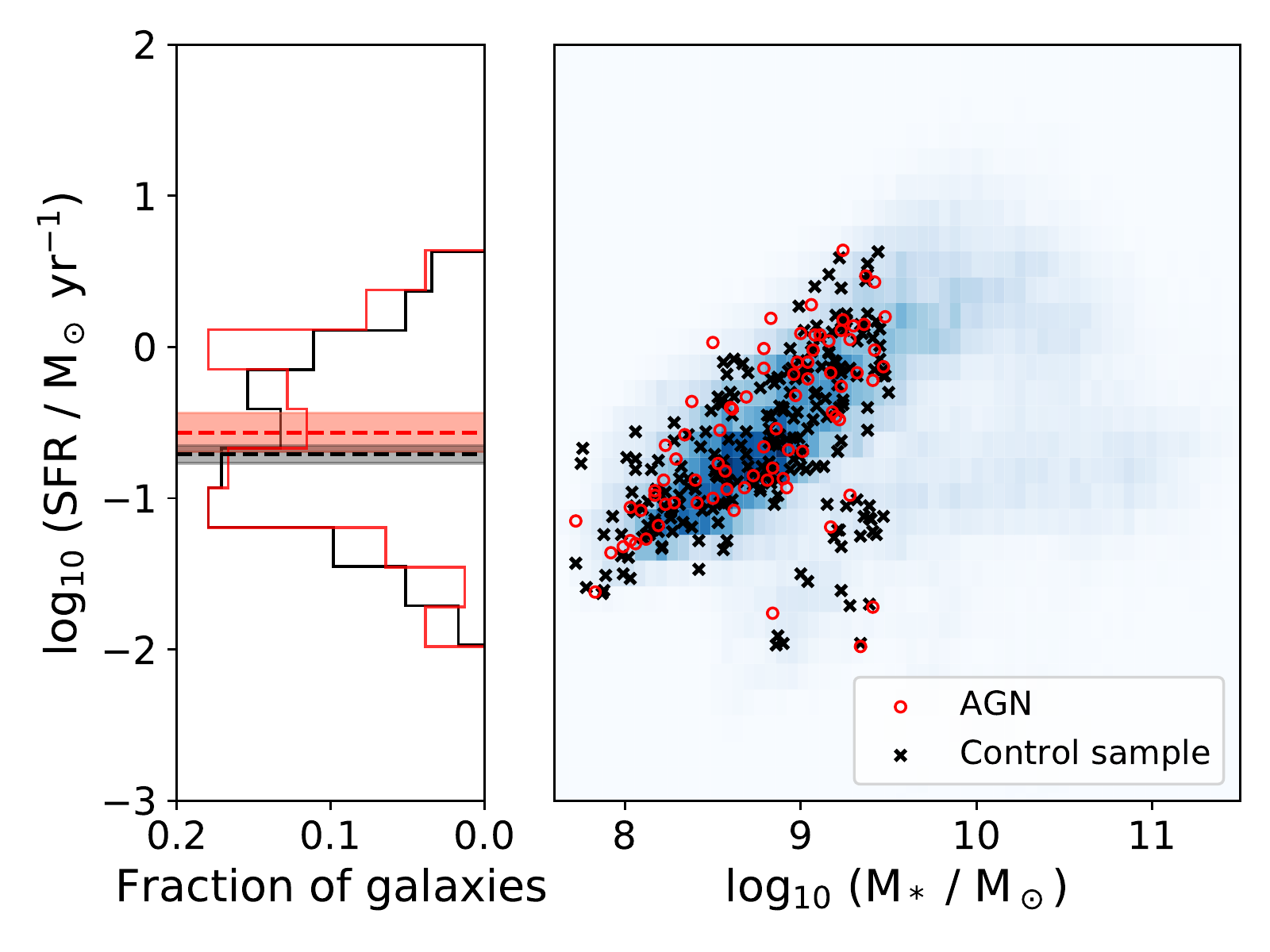}
    \includegraphics[width=\columnwidth]{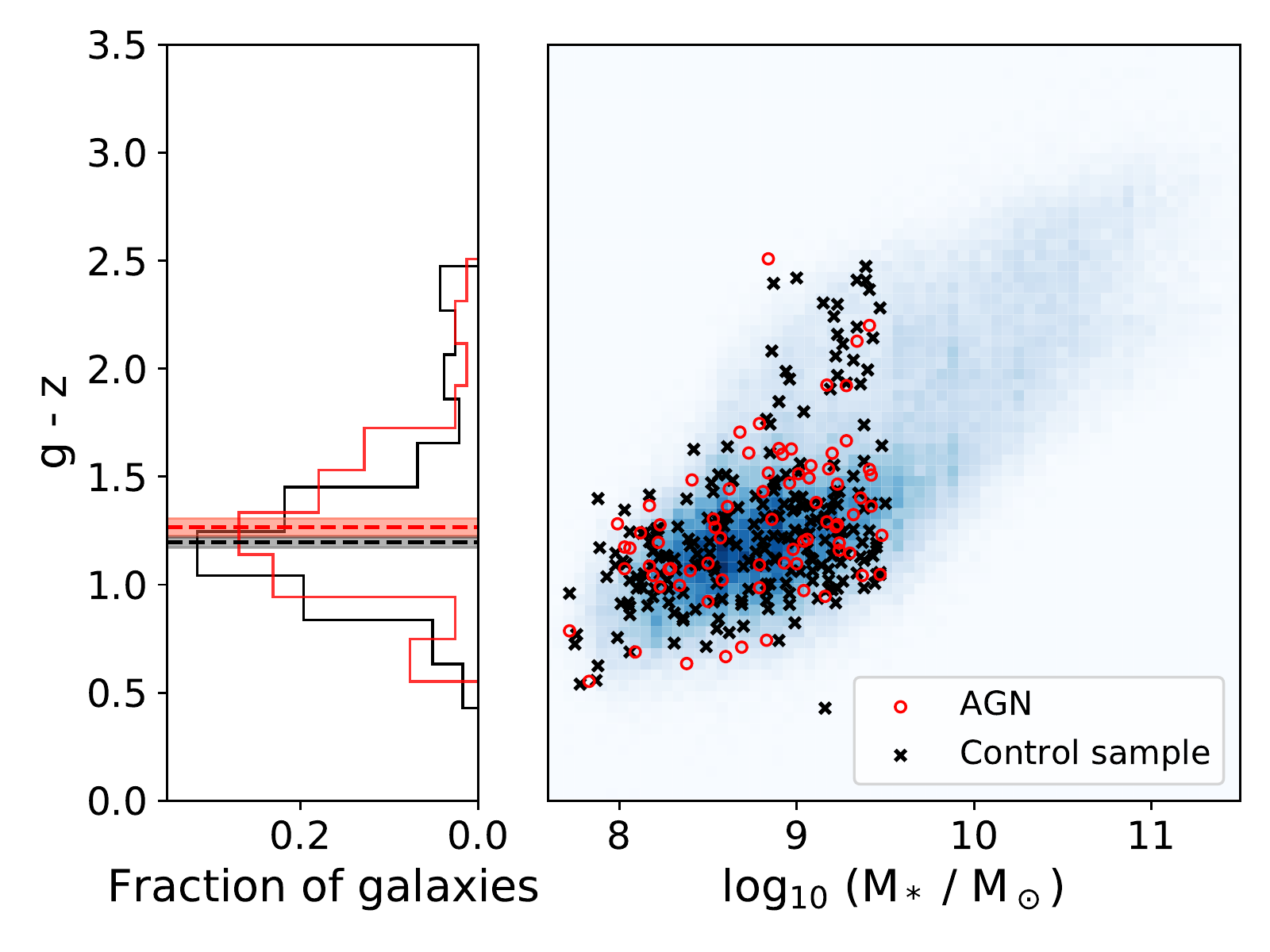}
    \caption{\textbf{Top:} SFR vs stellar mass for the AGN (red) and control sample (black). The heatmap shows the distribution of all galaxies in the HSC-SSP. The histograms show the distributions of SFRs for the AGN (red) and control sample (black). Medians and the corresponding bootstrapped uncertainties are shown using the vertical dashed lines and shaded regions respectively. \textbf{Bottom:} Same as top panel but for the observed $g-z$ colour (which roughly corresponds to rest-frame $u-r$ at the redshifts of our dwarf sample). There is no statistically significant difference between the distributions of the AGN and the controls in either the SFR or the observed $g-z$ colour.} 
    \label{fig:comparecontrols}
\end{figure}

In Figure \ref{fig:comparecontrols} we compare the SFR (top panel) and
observed $g-z$ colour (bottom panel) distributions of the AGN and
control samples. The observed $g-z$ colour (which roughly corresponds to
rest-frame $u-r$ at the redshifts of our dwarfs) traces the recent
star formation history. In all cases, the median values of the AGN and
control distributions are consistent within errors and a Mann Whitney
$U$ test between these distributions yields p-values greater than
0.05 in all cases, indicating that there are no statistically
significant differences in the distributions of both quantities.

The consistency in the SFR and colour suggests that the gas masses that are driving the star formation are similar in the AGN and the control samples. The AGN are, therefore, not gas-enriched compared to their control counterparts, either due to greater gas availability in their neighbourhoods (which appears consistent with the similarity in local environments described above) or because they are intrinsically more gas-rich.

\section{AGN properties - ages, jet powers, accretion rates and the plausibility of AGN feedback}
\label{sec:agn_properties}

\begin{figure*}
    \centering
    \includegraphics[width=0.49\textwidth]{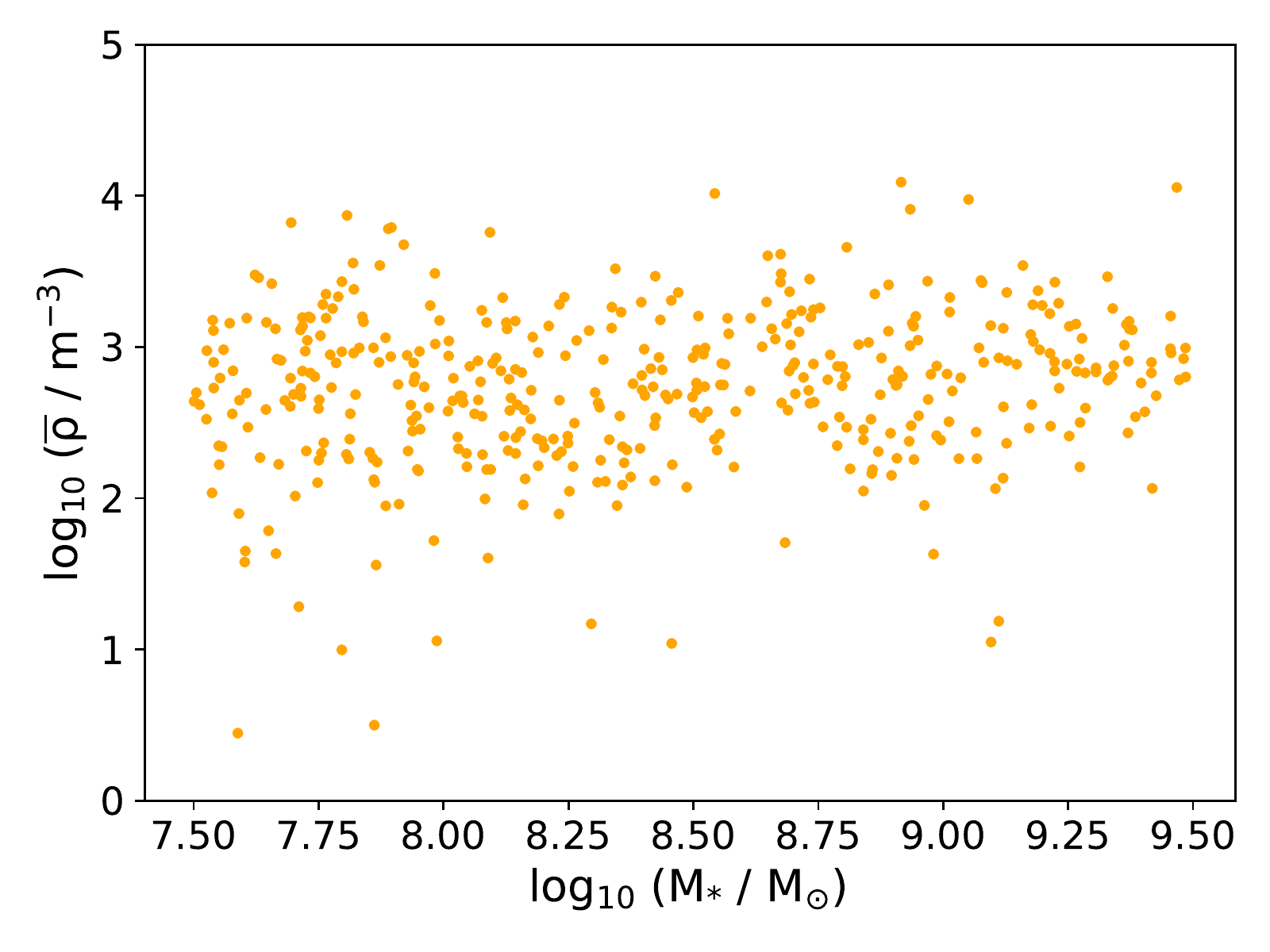}
    \includegraphics[width=0.49\textwidth]{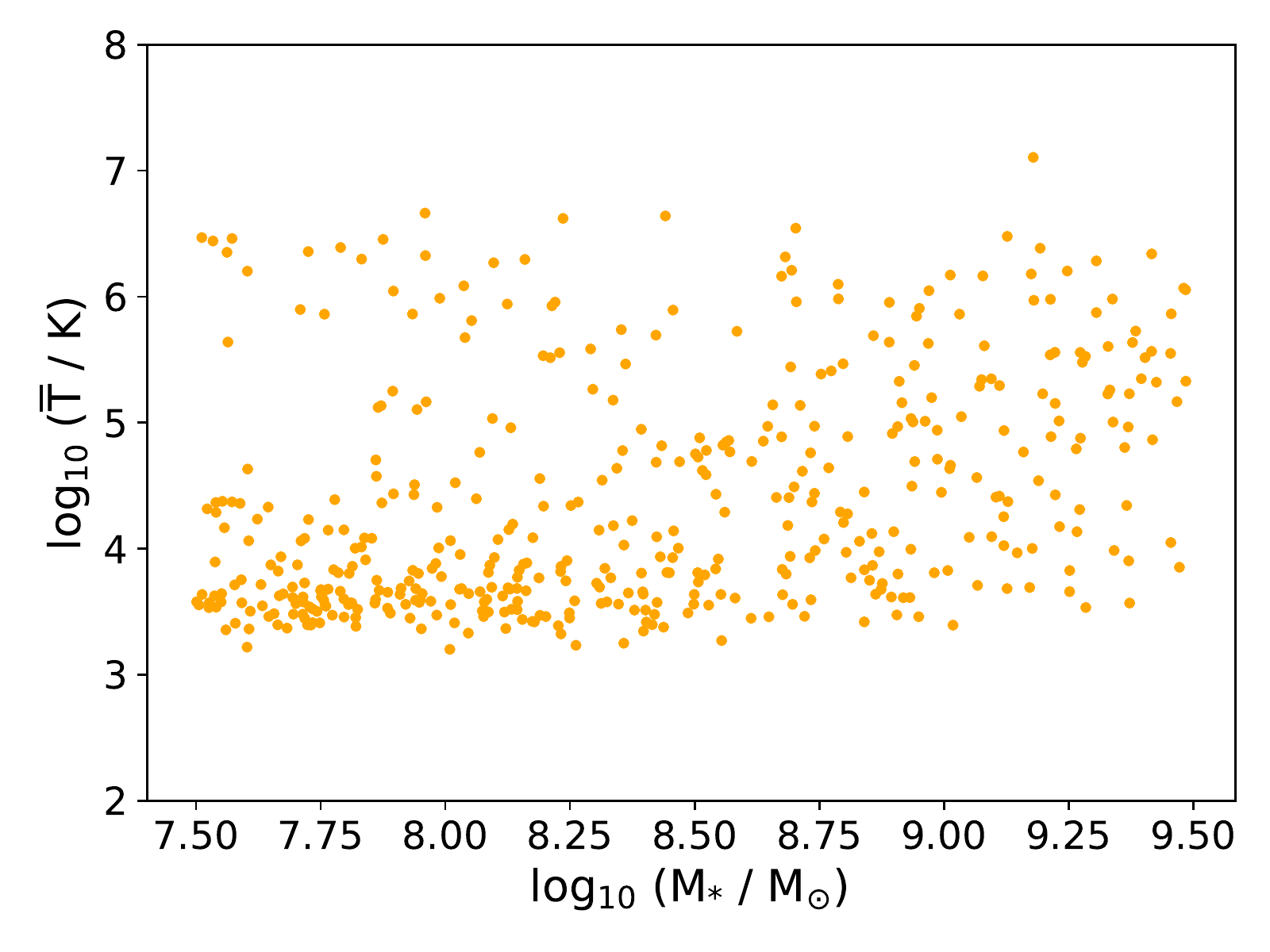}
    \includegraphics[width=0.49\textwidth]{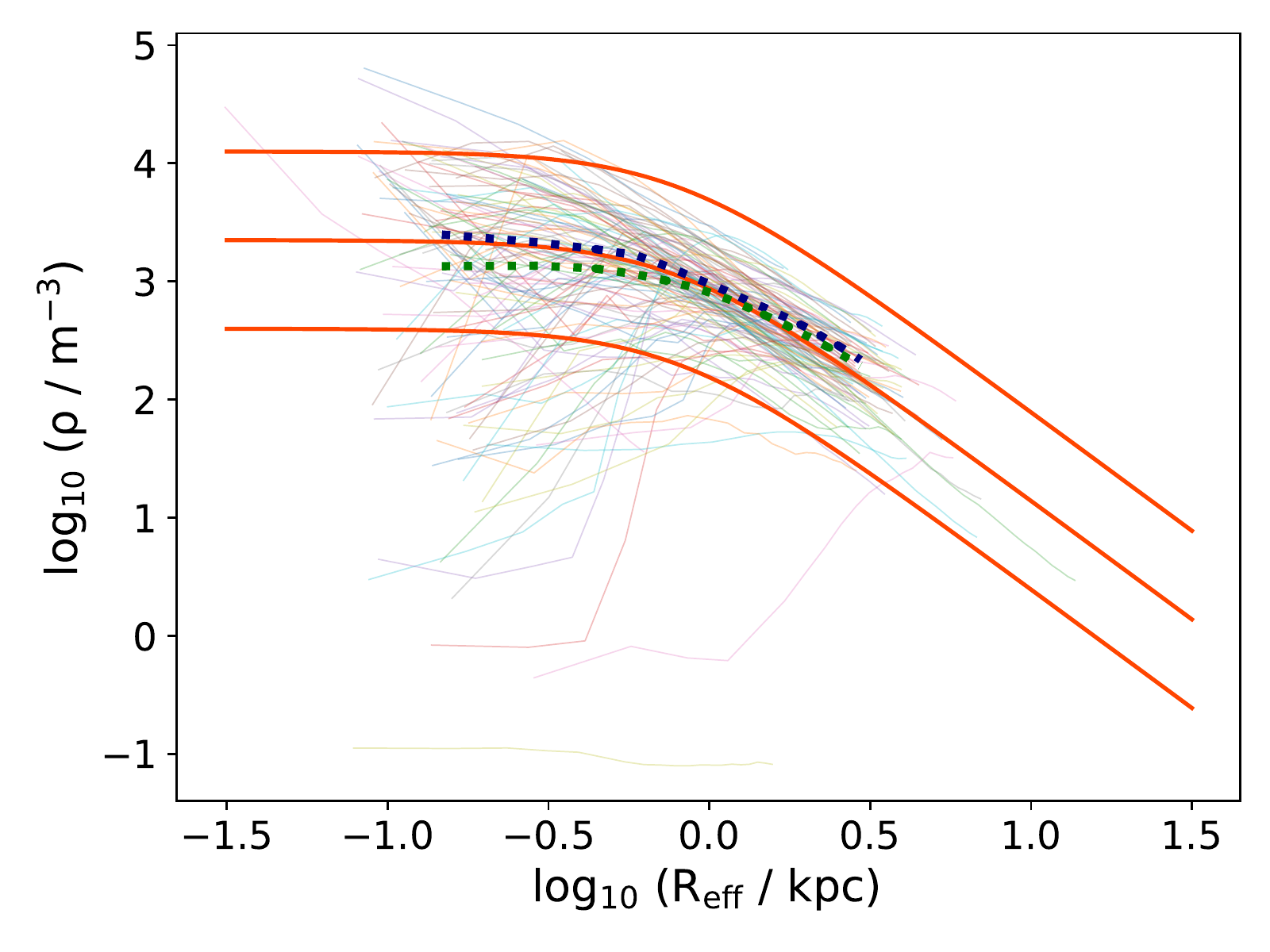}
    \includegraphics[width=0.49\textwidth]{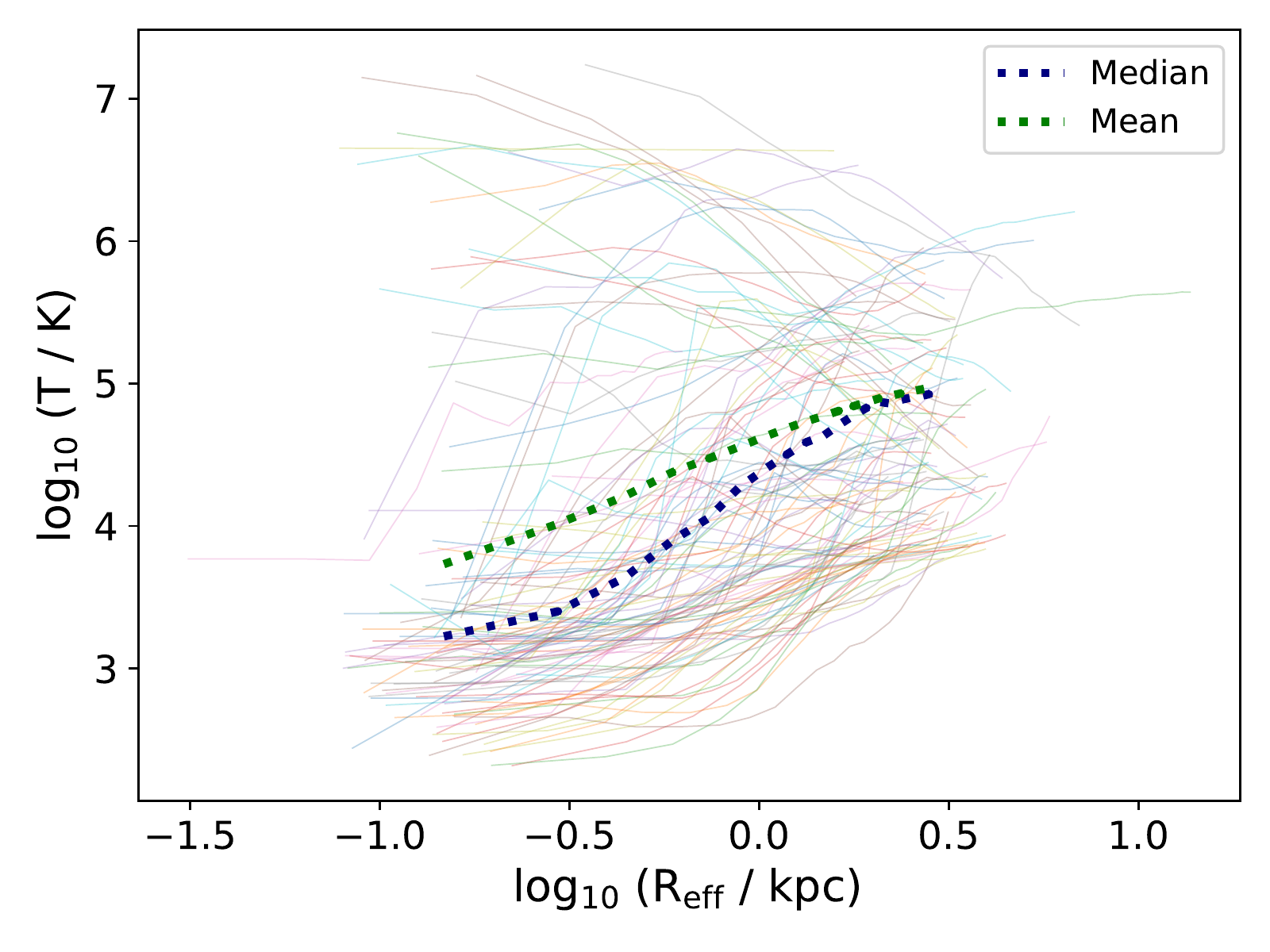}
    \caption{Mean central number densities and temperatures at 0.1 R$_{\text{eff}}$ (top row) and the profiles in these quantities (shown using hairlines in the bottom row) for dwarf galaxies from the \texttt{NewHorizon} cosmological simulation. The red solid lines indicate a range of King profiles \citep{King1966} that bracket the hairlines. The dotted lines in the bottom row indicate the geometric mean and median profiles (see legend).} 
    \label{fig:profiles}
\end{figure*}

We proceed by exploring the properties of the AGN in our dwarfs, by combining the gas conditions predicted in dwarf galaxies from the high-resolution \texttt{NewHorizon} (NH hereafter) cosmological hydrodynamical simulation \citep{Dubois2021} with a semi-analytical model for the evolution of radio galaxies \citep{Hardcastle2018}. Since we only have integrated radio luminosities for our dwarf AGN, and no information about the gas conditions in individual systems, we do not attempt to derive detailed properties for individual AGN in our sample. Instead, we use the ensemble of predicted initial conditions expected in dwarfs from the cosmological simulation to inform a Monte-Carlo suite of simulated radio AGN sources produced using the semi-analytical model. This enables us to statistically explore plausible ranges for properties like ages, jet powers and accretion rates for our AGN. We also compare the mechanical energy output of the simulated AGN in our Monte-Carlo suite to the expected binding energies of the dwarf gas reservoirs to study the plausibility of AGN feedback in these systems. 

In order to estimate the range of gas conditions which may exist in typical dwarf galaxies we use the predicted properties of dwarfs in the NH cosmological simulation at $z\sim0.3$. NH provides a spherical volume, with a radius of 10 comoving Mpc, centred on a region of average density within the larger (142 Mpc$^{3}$ volume) Horizon-AGN simulation \citep{Dubois2014,Kaviraj2017}. NH offers dark-matter and stellar mass resolutions of 10$^6$ and 10$^4$ M$_{\odot}$ respectively and a maximum spatial resolution of 34 pc. The gravitational force softening equals the local grid size. 
Galaxies in our stellar mass range of interest are, therefore, well-resolved in the simulation, with at least 5000 star particles representing individual objects. NH produces good agreement with key galaxy properties, over cosmic time, in both the dwarf and high mass regimes \citep{Dubois2021}.


In all NH dwarf galaxies, every gas cell within $1~R_{\rm eff}$ of the centre of each simulated galaxy is extracted at the maximum possible refinement (34 pc). We then calculate the average pressure, weighted by the volume of each gas cell, and the average density ($\sum m_{i}/(4/3 \pi R^{3})$) for all gas cells within radii ranging from $0.05~R_{\rm eff}$ out to a maximum of $1~R_{\rm eff}$ (where $m_{i}$ is the mass of an individual gas cell and $R_{\rm eff}$ is the effective radius). The average temperature within each radius bin is then calculated by assuming a single-phase gas. Figure \ref{fig:profiles} shows the mean densities and temperatures (top row) at 0.1 R$_{\text{eff}}$ and the profiles of these quantities in individual NH dwarfs (shown using hairlines in the bottom row). 

We then use the temperature and density profiles, derived as
described above, as input to a semi-analytical model for the evolution of radio galaxies \citep{Hardcastle2018}. The model solves differential
equations that model the shock front around radio galaxy lobes,
employing simplifying assumptions derived from 2D and 3D numerical
simulations. Based on existing observational constraints, it assumes a
light, electron-positron jet, solely radiating particles within the
lobes, and a magnetic field energy density that is one tenth that of
the radiating particles. For radio galaxies the model
reproduces a range of properties of observed radio sources (e.g. radio
luminosities, shocked shell temperatures, physical sizes) as well as
other inferences of observed trends between the jet power and the
radio luminosity \citep{Hardcastle2019,Mahatma2020}.


\begin{figure}
    \centering
    \includegraphics[width = \columnwidth]{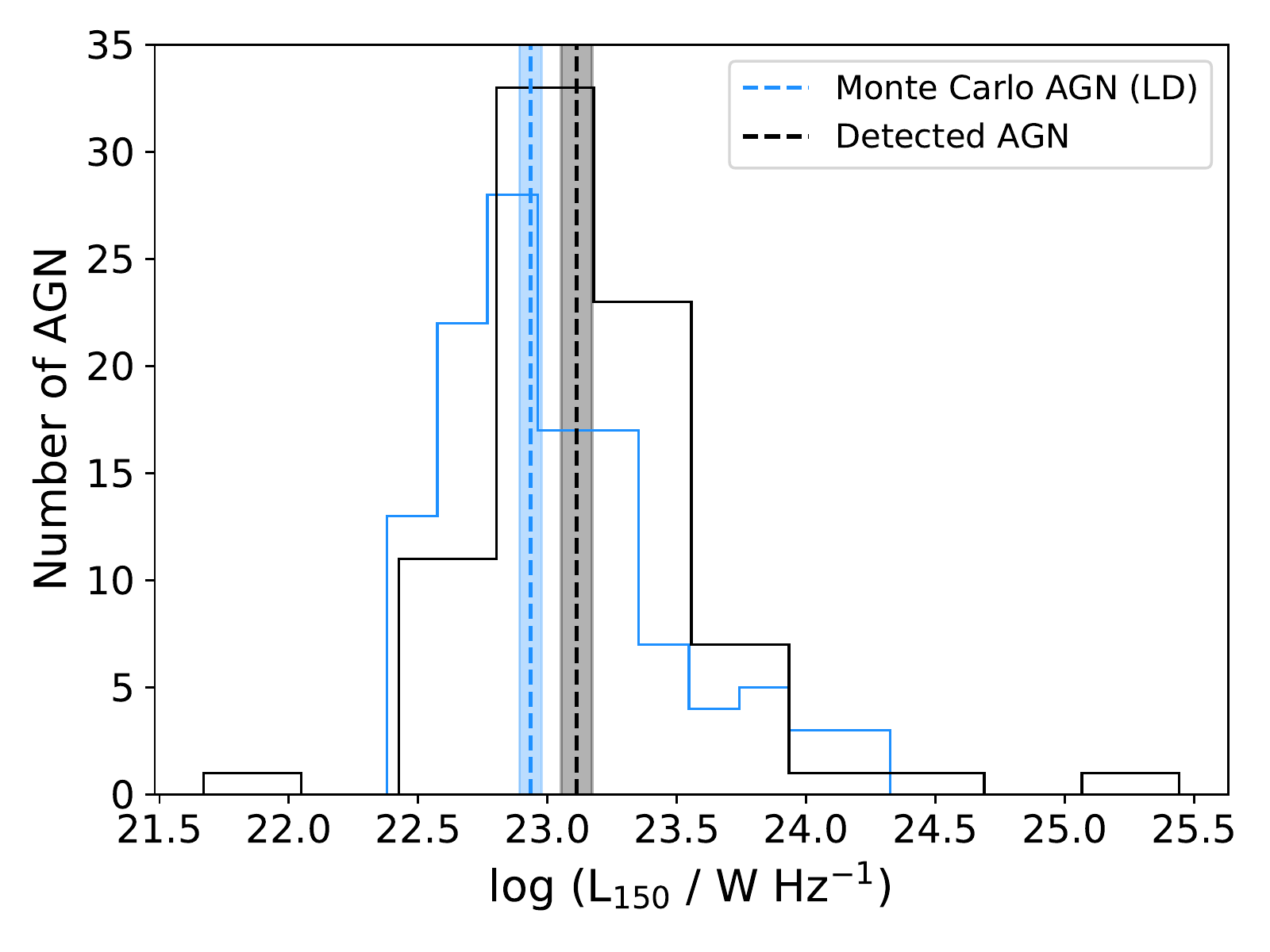}
    \caption{Distributions of the 150 MHz radio luminosities for the AGN (black) and the LOFAR-detectable (LD) sources in our Monte-Carlo suite of simulated radio AGN sources (blue). Median values and bootstrapped uncertainties are indicated by the dashed lines and shaded regions respectively. The distribution of radio luminosities of the LOFAR-detectable simulated sources are similar to those in our AGN.}
    \label{fig:radioplot}
\end{figure}

\begin{figure}
    \centering
    \includegraphics[width=\columnwidth]{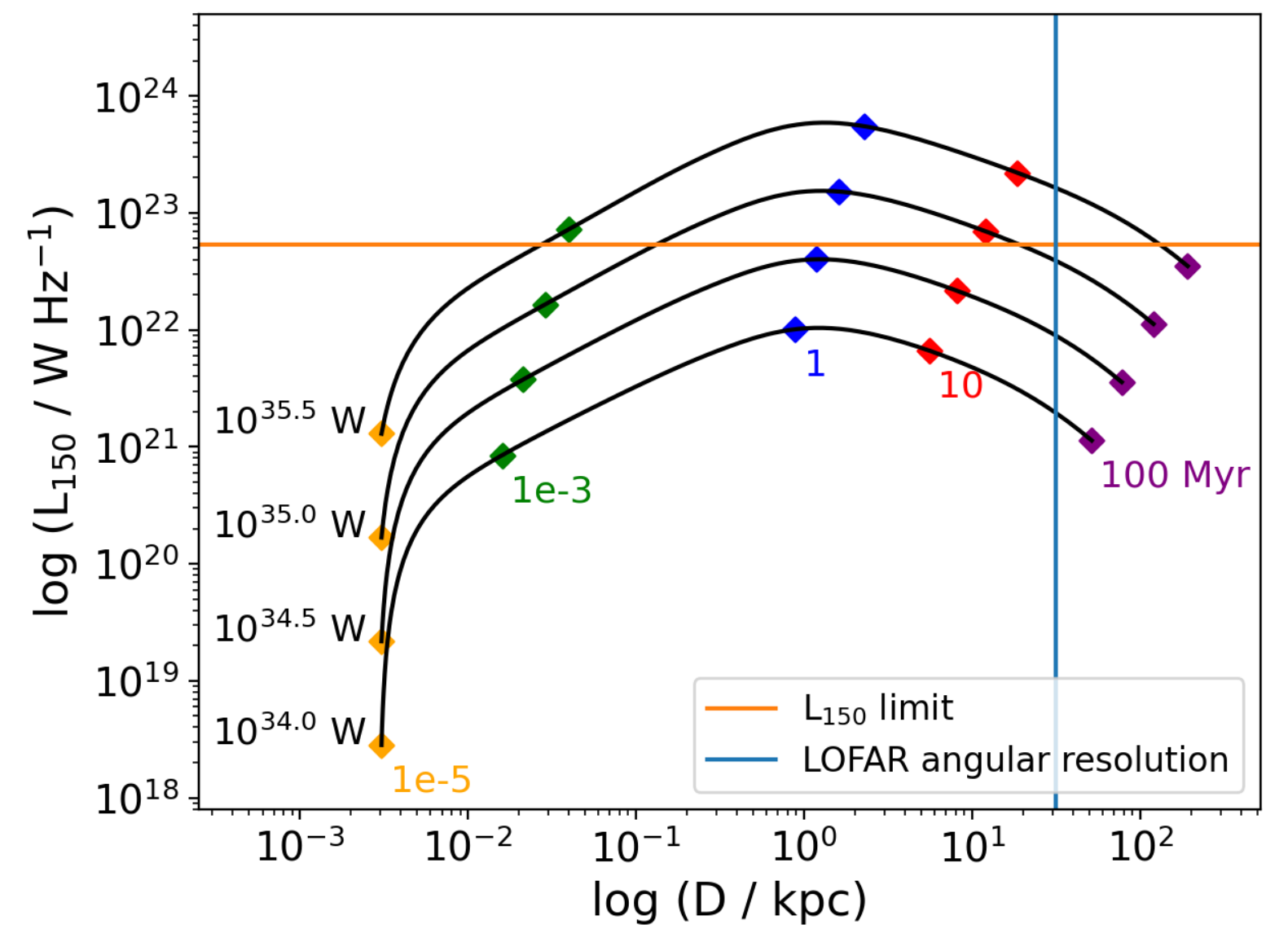}
    \caption{Examples of the evolution of typical LOFAR-detectable simulated sources with various jet powers. The time evolution, in Myr, is shown using the coloured points. The solid orange and blue lines indicate the LOFAR luminosity limit and the physical scale corresponding to the LOFAR angular resolution (6 arcsec) at the median redshift of our AGN ($z\sim 0.38$) respectively. Such typical evolutionary histories are expected to spend very little time ($<10$ Myr) brighter than the LOFAR detection limit after their onset. The AGN in this study are therefore young and close to their peak luminosities (where they are LOFAR-detectable) and, as a result, have relatively small physical sizes so that they are expected to be unresolved in the LOFAR images.}.  
    \label{fig:lengthlumtime}
\end{figure}

\begin{figure*}
    \centering
    \includegraphics[width=\textwidth]{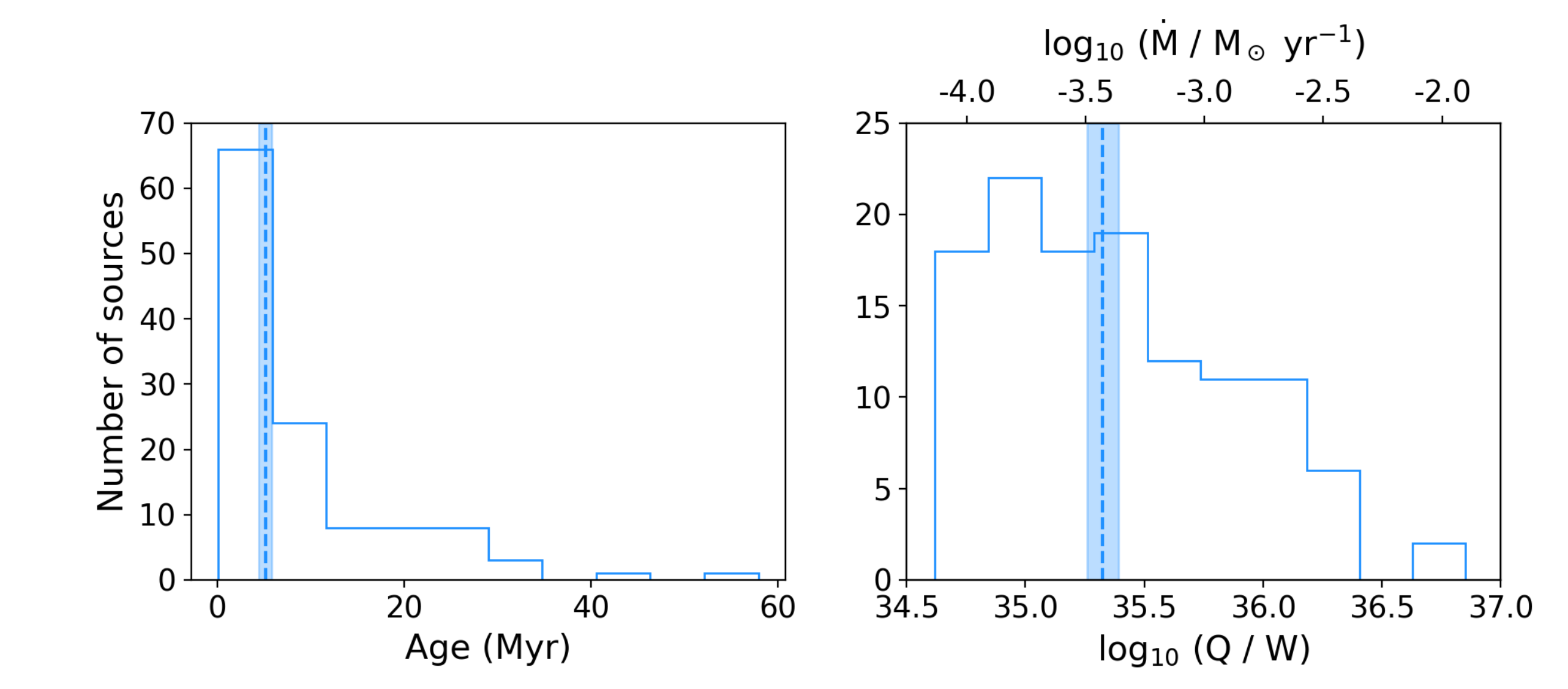}
    \caption{Distributions of ages (left) and jet powers (Q) and associated accretion rates ($\dot{\rm{M}}$; right) of the 119 LOFAR detectable sources in our Monte Carlo suite of simulated radio AGN sources. Median values and bootstrapped errors are shown using the dashed vertical lines and shaded regions respectively. The accretion rates, shown on the upper x-axis of the right-hand panel, are estimated using the jet powers (Q), assuming an efficiency ($\eta$) of 10 per cent, where $\dot{\rm{M}}=\rm{Q}/(\eta.\rm{c}^2)$.}
    \label{fig:radiomodels}
\end{figure*}

\begin{figure}
    \centering
    \includegraphics[width=\columnwidth]{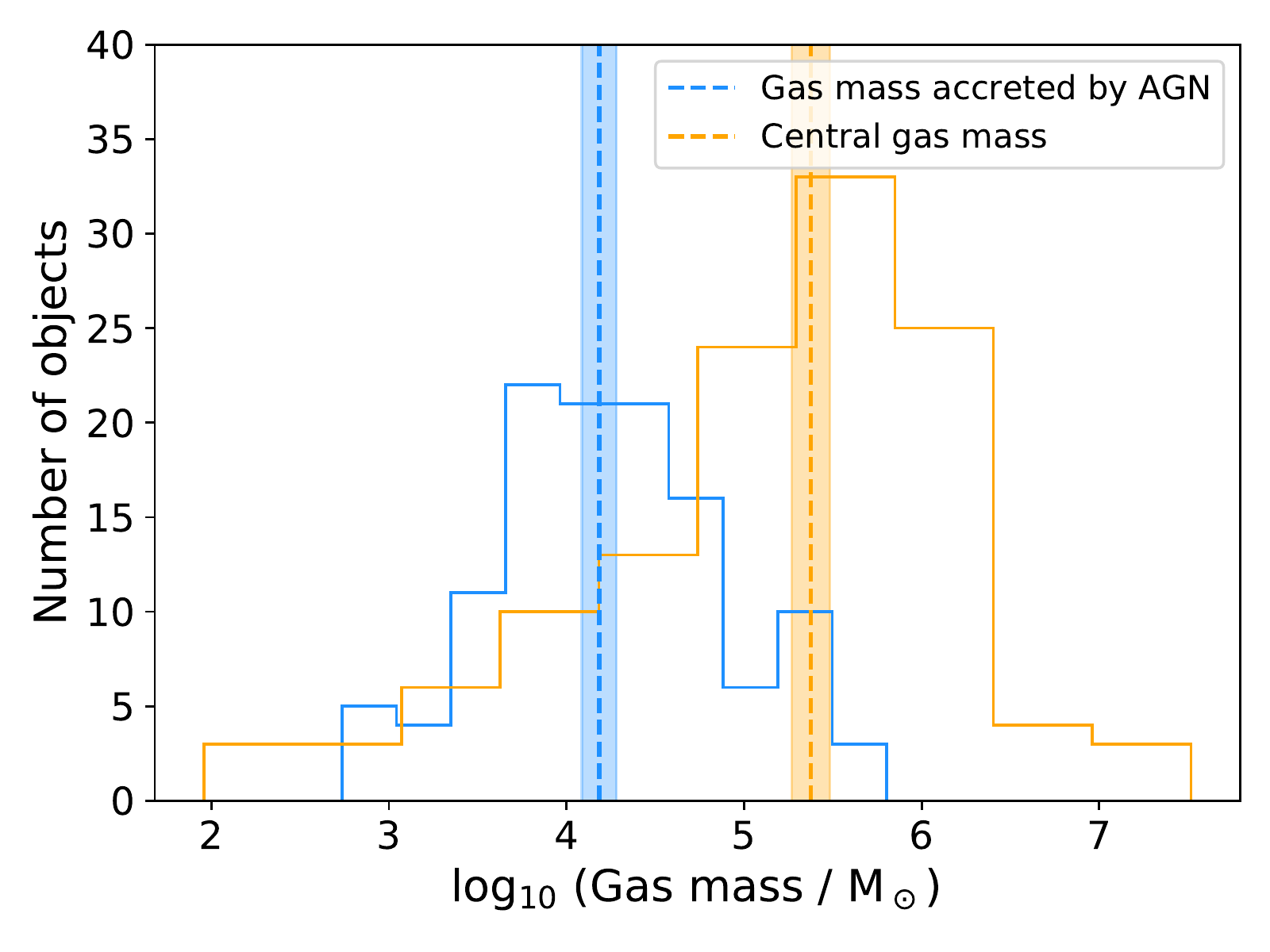}
    \includegraphics[width=\columnwidth]{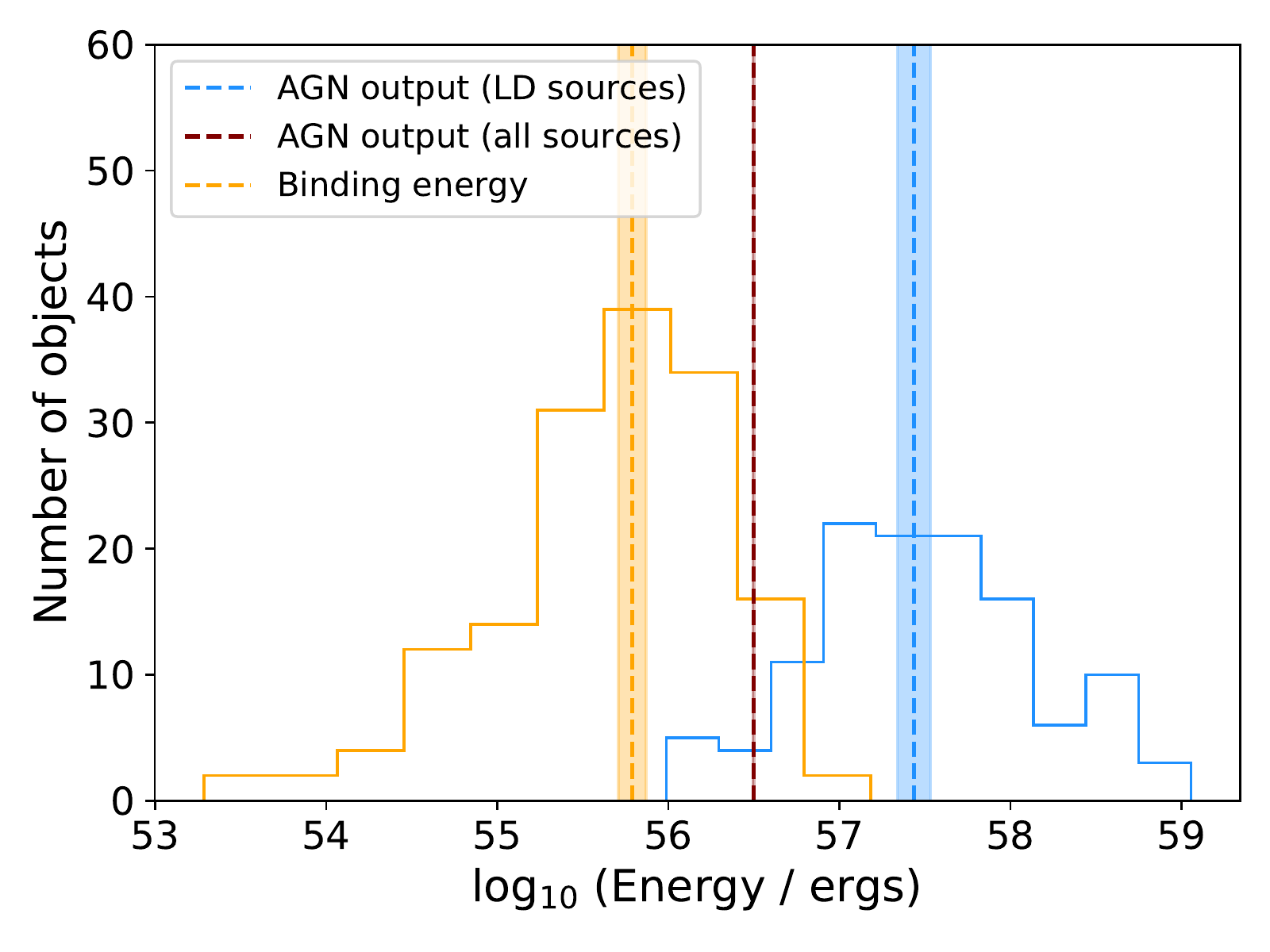}
    \caption{\textbf{Top:} Comparison of the central gas masses (within 0.05 R$_{\rm{eff}}$) in NH dwarfs (orange) to the total gas accreted onto the BH within the lifetime of the LOFAR-detectable simulated radio sources (blue). The NH dwarfs are mass matched to our AGN sample (middle panel in Figure \ref{fig:sampledistributions}). Median values and bootstrapped errors are shown using the dashed lines and shaded regions respectively. The central gas masses far exceed that required for AGN fuelling by more than an order of magnitude, indicating that there is an ample gas supply in the intrinsic gas reservoirs of dwarfs to trigger the observed AGN activity. \textbf{Bottom:} Comparison of the binding energies of the gas reservoirs, within 2 effective radii (orange), in NH dwarfs, to the mechanical energy outputted by the full Monte-Carlo suite of 10,000 simulated sources (red; the distribution is omitted for clarity) and the LOFAR-detectable (LD) subset (blue). Median values and bootstrapped errors are shown using the dashed vertical lines and shaded regions respectively. The median mechanical energy output of both the full suite of simulated sources and the LOFAR-detectable subset is higher than the median binding energy of the gas reservoirs by $\sim 0.8$ and $\sim 2$ dex respectively. The mechanical energy output of AGN in such systems is therefore likely to be able to perform significant amounts of negative feedback as is the case in massive galaxies.}
    \label{fig:energies}
\end{figure}

We use this modelling to determine the evolution of a Monte-Carlo
suite of 10,000 simulated radio AGN sources in the types of
environments expected in dwarf galaxies, in order to provide a broad
statistical exploration of the parameter space that is likely to give
rise to the AGN that we see in the dwarf galaxies in our observed
sample. The procedure follows that outlined in Section 4 of
\cite{Hardcastle2018}. For these simulated sources, we draw redshifts
from a uniform distribution in the range $0.3 < z < 0.5$ (the redshift
range within which most of our dwarfs lie, see Figure
\ref{fig:sampledistributions}), jet powers from a power-law
distribution in the range $10^{34} < Q < 10^{37}$ W, with $p(Q) \propto
Q^{-1}$ (which produces a good match to the observed radio luminosity function of more powerful radio-loud AGN, see \citealt{Hardcastle2019}) and temperatures from a power-law distribution in the range $3500 < T < 3 \times 10^{6}$ K with $p(T) \propto T^{-1}$ to roughly match the results from NH. 

All sources are
assumed to have an isothermal King profile atmosphere with a core
radius of 0.8 kpc, as this gives a reasonable match to the density
profiles from NH (Figure \ref{fig:profiles}), and central densities are
taken from a log-uniform distribution in the range shown in the bottom
left panel of Figure \ref{fig:profiles}, i.e.
log$_{10}(n_0/\mathrm{m}^{-3}) = 3.35 \pm 0.75$. The source lifetimes
  are drawn from a uniform distribution from 0 to 100 Myr and the AGN
  episode start time is drawn from a uniform distribution between 120
  and 0 Myr before the time of simulated observation: since a remnant
  radio AGN in which the jet has switched off will fade below the
  detection threshold in less than 20 Myr, this ensures that the
  simulations cover all observable phases of a dwarf radio AGN's
  evolution. Each source is also assigned a random angle to the line
  of sight drawn from a distribution with $p(\theta) = \sin(\theta)$. Once these simulations are complete, we simulate
  observations of the source population at the LOFAR observing
  frequency, including the effects of spectral ageing but also the
  flux and surface-brightness limits of the LOFAR data, in order to
  characterise the population of the simulated objects that would be
  detectable by LOFAR.

The subset of LOFAR-detectable sources in our Monte-Carlo suite shows
a remarkably similar distribution of radio luminosities to that of the AGN in our dwarf galaxies (Figure \ref{fig:radioplot}). Figure \ref{fig:lengthlumtime} shows examples of the radio luminosity and source-length evolution of typical model sources, with various jet powers. The time evolution, in Myr, is shown using the coloured points. The solid orange and blue lines indicate the LOFAR luminosity limit and the physical scale corresponding to the LOFAR angular resolution (6 arcsec) at the median redshift of our dwarf AGN ($z\sim 0.38$) respectively. This figure indicates that sources with such typical evolutionary histories are expected to spend very little time (less than a few tens of Myr) at radio luminosities brighter than the LOFAR detection limit after their onset. The AGN in this study are therefore ones which are young and close to their peak luminosities (where they are LOFAR detectable) and, as a result, have small physical sizes, so that they are not resolved in the LOFAR images. These properties appear consistent with the fact that the colours of the AGN and controls are similar (as would be expected if we are preferentially detecting young AGN, which have not had a chance to significantly impact their gas reservoirs) and the fact that we do not see any extended sources in our sample in our visual inspection\footnote{It is worth noting here that, since our AGN are preferentially young and their radio luminosities are likely to decay rapidly, it is plausible that a fraction of the dwarfs in the control sample do in fact host faint AGN which fall below the detection threshold of LOFAR.}.

In the left and right-hand panels of Figure \ref{fig:radiomodels}, we
present the distributions of ages and jet powers of the 119
LOFAR-detectable sources in our Monte Carlo simulation respectively. The median
values of the jet powers and ages are 5.17 Myr and $10^{35.3}$ W. The
16 (84) percentile values are 1.03 (18) Myr and $10^{34.9}$
($10^{35.9}$) W respectively. Given the good agreement between
the observed and simulated properties of this population, and if the
assumptions implicit in our modelling are correct, these jet powers
and ages should be approximately representative of the true observed
population.


We use the jet powers ($\rm{Q}$) to estimate accretion rates
($\dot{\rm{M}}$) assuming an efficiency ($\eta$) of 10 per cent\footnote{This is simply in analogy with the radiative efficiency generically used to estimate accretion rate from bolometric luminosity. However, jet efficiency can be very different from this value, by up to an order of magnitude \citep{Russell2013,Nemmen2015}} \citep[e.g.][]{Fabian2012}, where
$\dot{\rm{M}}=\rm{Q}/(\eta.\rm{c}^2)$. These accretion rates are shown
on the upper $x$-axis in the top panel of this figure. The median
accretion rate is $\sim 10^{-3.43}$ M$_{\odot}$ yr$^{-1}$ and the 16
and 84 percentile values are $10^{-3.89}$ and $10^{-2.81}$
M$_{\odot}$ yr$^{-1}$ respectively. It is worth noting that, given these accretion rates, the gas mass accreted by the AGN during its lifetime is small compared to that already available in the intrinsic gas reservoirs of nearby dwarf galaxies. 

To demonstrate this point, we construct a sample of NH dwarfs that are mass-matched to our observed AGN sample, by selecting, for every observed AGN, two random NH dwarfs that have stellar masses within 0.05 M$_{\odot}$ of the AGN in question. The top panel of Figure \ref{fig:energies} indicates that, in the very central regions of these NH dwarfs (i.e. within the inner 5 per cent of their effective radii), where the gas is likely to be more susceptible to accretion by the BH, the median gas masses (orange) are more than an order of magnitude larger than what is needed to fuel the LOFAR-detectable simulated sources from our semi-analytical analysis (blue). 
In other words, an abundant gas supply likely exists within the existing gas reservoirs of nearby dwarfs to drive the observed AGN activity. 


We complete our study by comparing the expected binding energy of gas
reservoirs in dwarfs to the mechanical energy output of the simulated
radio AGN sources. In the bottom panel of Figure \ref{fig:energies}, we compare the
binding energies of the gas reservoirs of the same mass-matched NH dwarfs
(orange) to the mechanical energy output of the LOFAR-detectable
simulated sources (light blue). The binding energies are estimated by
calculating the gravitational potential of a uniform sphere, using the
total mass inside two effective radii. The mechanical energy output
by the radio sources is calculated by multiplying the jet power $Q$ (which
is assumed to be constant during the active phase in these semi-analytical models) by the lifetime of the AGN. Median values are shown using the dashed lines while bootstrapped uncertainties are shown using the shaded regions. We also show the median value of the mechanical energy output of all 10,000 sources in our Monte-Carlo suite (dark red). We omit the distribution for clarity. 

While the LOFAR-detectable simulated sources are consistent with our observed AGN population in terms of their radio luminosities, it is not clear whether the `LOFAR-undetected' sources have counterparts in the real Universe. Nevertheless, given the broad parameter space traced by the Monte-Carlo suite, comparing the mechanical energy output of the LOFAR-undetected sources is useful for gauging the general plausibility of feedback in radio AGN within dwarfs, rather than just in the youngest systems. The median mechanical energy output of both the full suite of 10,000 simulated AGN sources (10$^{56.5}$ erg) and the subset that is LOFAR-detectable (10$^{57.4}$ erg) are higher than the median binding energy (10$^{55.7}$ erg) by $\sim 0.8$ and $\sim 2$ dex respectively. This implies that removal and/or displacement of significant fractions of the gas reservoir via energetic feedback from the AGN is plausible in dwarfs (as is the case in massive galaxies).

In summary, our analysis indicates that special circumstances (e.g. particular environments, the presence of interactions or higher gas fractions) are not required to trigger the AGN we observe in our dwarf galaxies. Furthermore, an ample gas supply exists within the central regions of the intrinsic gas reservoirs of dwarfs to drive the observed AGN activity. Taken together, this suggests that AGN triggering in this regime is likely to be stochastic in nature. Since there is no reason to believe that the epochs probed in this study are special in terms of triggering AGN in dwarfs, this could be a rather common phenomenon. The young AGN we find in our study are not inconsistent with the short-lived AGN episodes at high accretion rates hypothesised in current theoretical studies, where BHs are unable to grow effectively in dwarfs \citep{Dubois2015,Habouzit17}. Our inability to observe fainter long-lived AGN at the depth of our LOFAR images makes it difficult to constrain the true lifetime distribution of AGN in dwarfs and put constraints on fuelling and growth of BHs in such galaxies. Observing long-lived AGN, accreting at high fractions of the Eddington rate, would be needed in order to show that BHs in dwarf galaxies can grow efficiently.

It is worth noting that the low AGN fractions reported in many previous studies (e.g. those noted in the introduction) are likely to be driven by the fact that the low-mass BHs in dwarfs are only detectable in existing surveys when they are close to their peak accretion rates, rather than due to a general paucity of AGN activity in dwarf galaxies. It is also for this reason that estimating an AGN fraction in our study is not meaningful. Finally, the potential mechanical energy output of the AGN in our dwarfs are likely to exceed the typical binding energy of their gas reservoirs, making AGN feedback a plausible process in these systems.

\section{Summary}
\label{sec:summary}

Dwarf galaxies dominate the galaxy number density, making them essential to a complete understanding of galaxy evolution. While AGN are thought to play an important role in influencing the evolution of massive galaxies (e.g. via negative feedback which regulates their stellar mass growth), the role of AGN in the dwarf regime is a key question that remains largely unexplored. BH growth within dwarfs in current cosmological simulations is stunted, both due to supernova feedback blowing away ambient gas around the BHs and also because these BHs tend to wander in regions of low gas density \citep[e.g.][]{Dubois2021}. However, it is not clear whether this behaviour is a result of the low spatial resolution of these simulations (and thus the inability to properly resolve the structures close to the BHs themselves) or whether this is an accurate representation of BHs in dwarf galaxies in the real Universe. Observational studies which can offer constraints on such simulations are therefore valuable. 

Empirical studies of accreting BHs in dwarf galaxies have been hampered both by the depths of past surveys and by strong selection biases (e.g. dwarfs in shallow surveys, like the SDSS, requiring anomalously high SFRs to be detectable in the first place). 
Nevertheless, recent observational work has provided a wealth of evidence, not just for the existence of BHs in dwarf galaxies, but also for sustained periods of accretion in these systems (both of which were points of significant debate even a decade ago). However, there has remained a lack of insight into the factors and processes that trigger BHs in the dwarf regime, the properties of the AGN themselves (e.g. jet powers and accretion rates) and how the energetics of the AGN compare to the binding energies of the gas reservoirs, which is critical for determining the plausibility of AGN feedback. 


In this study, we have combined deep optical and radio data in the ELAIS-NI field, from the HSC-SSP and LOFAR respectively, to study 78 nearby ($z<0.5$) dwarf galaxies that host radio AGN. We have then compared the properties of these AGN to a control sample that is matched in stellar mass and redshift. Our analysis of the characteristics of the AGN has used the predicted gas conditions in dwarfs from a cosmological hydrodynamical simulation as input into a Monte-Carlo suite of simulated radio sources from a semi-analytical model for radio-galaxy evolution. We have used this to explore the AGN properties (e.g. ages, jet powers and accretion rates) and compared the expected binding energies of dwarf gas reservoirs to the potential mechanical energy output of the AGN, in order to explore the plausibility of AGN feedback. Our main conclusions are as follows:



\begin{itemize}

\item The local environments of our AGN are similar to that of their control counterparts, with no evidence that the AGN are preferentially closer to massive galaxies. The triggering of these AGN is therefore unlikely to be driven by environmental factors (e.g. higher rates of gas accretion due to a denser local environment or triggering of the internal gas reservoirs of the dwarf AGN hosts due to stronger tidal fields). 

\item The fraction of AGN with tidal features, which are indicative of recent interactions, is relatively low ($\sim 6$ per cent) and consistent with that in their control counterparts. This suggests that neither mergers with lower mass galaxies nor fly-bys with massive galaxies are the principal triggering process in dwarfs which host AGN. 


\item At the redshifts of our AGN, the LOFAR detection limit is close to the peak luminosities of the subset of sources in our Monte-Carlo simulation that are LOFAR-detectable. These peak luminosities also decay rapidly, to values fainter than the detection limit, over timescales of a few tens of Myr. This indicates that the AGN in our study are preferentially young (the median age is only $\sim 5$ Myr).

\item The distributions of SFRs and observed $g-z$ colours of the AGN and control samples are consistent with each other. The observed $g-z$ colour roughly traces rest-frame $u-r$ at our redshifts of interest. Note that, since our AGN are young, the SFRs and colours trace the gas richness of the dwarf host just before the AGN switched on. The consistency in the SFRs and colours suggests that the gas masses that are driving the star formation are similar in the AGN and controls. The AGN are, therefore, unlikely to be gas-enriched compared to their control counterparts, either due to greater gas availability in their neighbourhoods (which appears consistent with the similarity in local environments described above) or because they are intrinsically more gas-rich.  


\item In the subset of the LOFAR-detectable sources in our Monte-Carlo suite (which exhibits a similar distribution of radio luminosities as our observed AGN), the median values of the ages, jet powers, and accretion rates (assuming a canonical 10 per cent efficiency) are $\sim 5$ Myr, $\sim 10^{35}$ W and $\sim 10^{-3.4}$ M$_{\odot}$ yr$^{-1}$ respectively. 


\item The gas masses consumed by these simulated sources are more than an order of
  magnitude lower than the predicted gas masses in the central regions
  of their dwarf hosts. There is, therefore, likely to be an ample internal gas supply to trigger the (relatively low) accretion rates required to drive the AGN seen in our sample. 

\item A comparison of the expected binding energies of dwarf gas reservoirs to the mechanical energy input of simulated sources, in both the full Monte-Carlo suite and the subset that is LOFAR-detectable, indicates that the binding energies are at least an order of magnitude lower than the mechanical energy available from the AGN. Negative AGN feedback is, therefore, as plausible in dwarfs as it is in massive galaxies. 

\item Our analysis shows that special circumstances (e.g. particular
  environments, the presence of interactions or higher gas fractions)
  are not necessary to trigger the AGN in our dwarf galaxies and an
  ample gas supply likely exists within their intrinsic gas reservoirs
  to drive the AGN activity. This suggests that AGN
  triggering in this regime is likely to be stochastic in nature.
  Furthermore, since there is no reason to believe that the epochs
  sampled in this study are special in terms of AGN activity in
  dwarfs, this could be a rather common phenomenon. 
  It is likely that the low AGN fractions reported in some previous studies are driven by the fact that the low-mass BHs in dwarfs are only detectable in existing surveys when they are close to their peak accretion rates, rather than due to a general dearth of AGN activity in dwarf galaxies. Indeed, it is worth recalling here that in very nearby dwarfs ($cz<10,000$ km s$^{-1}$), the fraction of systems that exhibit AGN signatures, can be high \citep[e.g. $\sim 80$ per cent,][]{Dickey2019}. 



\end{itemize}

While the combination of the HSC-SSP and LOFAR has enabled us to the study young AGN in dwarf galaxies, which are close to their peak radio luminosities, forthcoming surveys will enable us to gain deeper insights into the broader population of AGN in dwarf galaxies. LSST, which will offer similar depth to the HSC-SSP Deep layer over around half the sky, will image millions of dwarf galaxies in six optical filters. When combined with radio data from the Square Kilometre Array \citep[SKA, e.g.][]{Dewdney2009}, this will enable uniquely detailed studies of radio AGN in the dwarf regime. Moreover, dwarf galaxies that host an AGN could be studied using LSST data alone, by exploiting their optical variability. The depth of these forthcoming datasets will permit the detection of AGN in lower mass galaxies, with a much broader range of accretion rates and ages, than is possible using current instrumentation.

To conclude, our study suggests that AGN potentially play as significant a role in the evolution of dwarfs as they do in massive galaxies. The implementation of AGN physics in the dwarf regime should be explored in more detail in galaxy formation models. Confrontation of these models with forthcoming surveys, like the LSST and SKA, will be critical for advancing our understanding of how dwarf galaxies form over cosmic time and the role that AGN play in their evolution. 


\section*{Acknowledgements}

We are grateful to the anonymous referee for several suggestions that improved the quality of the original manuscript. FD, SK, MH and RAJ acknowledge support from the STFC [ST/V506709/1,  ST/S00615X/1, ST/V000624/1, ST/R504786/1]. KM has been supported by the National Science Centre (UMO-2018/30/E/ST9/00082). RAJ acknowledges support from the Yonsei University Research Fund (Yonsei Frontier Lab. Young Researcher Supporting Program) of 2021 and from the Korean National Research Foundation (NRF-2020R1A2C3003769). 

Some of the numerical work made use of the DiRAC Data Intensive service at Leicester, operated by the University of Leicester IT Services, which forms part of the STFC DiRAC HPC Facility (www.dirac.ac.uk). The equipment was funded by BEIS capital funding via STFC capital grants ST/K000373/1 and ST/R002363/1 and STFC DiRAC Operations grant ST/R001014/1. DiRAC is part of the National e-Infrastructure. This research has used the DiRAC facility, jointly funded by the STFC and the Large Facilities Capital Fund of BIS, and has been partially supported by grant Segal ANR- 19-CE31-0017 of the French ANR. This work was granted access to the HPC resources of CINES under the allocations 2013047012, 2014047012 and 2015047012 made by GENCI and was granted access to the high-performance computing resources of CINES under the allocations c2016047637 and A0020407637 from GENCI, and KISTI (KSC-2017-G2-0003). Large data transfer was supported by KREONET, which is managed and operated by KISTI. This work has made use of the Horizon cluster on which the simulation was post-processed, hosted by the Institut d'Astrophysique de Paris. We thank Stephane Rouberol for running it smoothly for us. 

The Hyper Suprime-Cam (HSC) collaboration includes the astronomical communities of Japan and Taiwan, and Princeton University. The HSC instrumentation and software were developed by the National Astronomical Observatory of Japan (NAOJ), the Kavli Institute for the Physics and Mathematics of the Universe (Kavli IPMU), the University of Tokyo, the High Energy Accelerator Research Organization (KEK), the Academia Sinica Institute for Astronomy and Astrophysics in Taiwan (ASIAA), and Princeton University. Funding was contributed by the FIRST program from Japanese Cabinet Office, the Ministry of Education, Culture, Sports, Science and Technology (MEXT), the Japan Society for the Promotion of Science (JSPS), Japan Science and Technology Agency (JST), the Toray Science Foundation, NAOJ, Kavli IPMU, KEK, ASIAA, and Princeton University. 

This paper makes use of software developed for the Large Synoptic Survey Telescope. We thank the LSST Project for making their code available as free software at \url{http://dm.lsst.org}.

The Pan-STARRS1 Surveys (PS1) have been made possible through contributions of the Institute for Astronomy, the University of Hawaii, the Pan-STARRS Project Office, the Max-Planck Society and its participating institutes, the Max Planck Institute for Astronomy, Heidelberg and the Max Planck Institute for Extraterrestrial Physics, Garching, The Johns Hopkins University, Durham University, the University of Edinburgh, Queen’s University Belfast, the Harvard-Smithsonian Center for Astrophysics, the Las Cumbres Observatory Global Telescope Network Incorporated, the National Central University of Taiwan, the Space Telescope Science Institute, the National Aeronautics and Space Administration under Grant No. NNX08AR22G issued through the Planetary Science Division of the NASA Science Mission Directorate, the National Science Foundation under Grant No. AST-1238877, the University of Maryland, and Eotvos Lorand University (ELTE) and the Los Alamos National Laboratory.

Based in part on data collected at the Subaru Telescope and retrieved from the HSC data archive system, which is operated by Subaru Telescope and Astronomy Data Center at National Astronomical Observatory of Japan. 

LOFAR, the Low Frequency Array designed and constructed by ASTRON, has
facilities in several countries, which are owned by various parties
(each with their own funding sources), and are collectively operated
by the International LOFAR Telescope (ILT) foundation under a joint
scientific policy. The ILT resources have benefited from the
following recent major funding sources: CNRS-INSU, Observatoire de
Paris and Universit\'e d'Orl\'eans, France; BMBF, MIWF-NRW, MPG, Germany;
Science Foundation Ireland (SFI), Department of Business, Enterprise
and Innovation (DBEI), Ireland; NWO, The Netherlands; the Science and
Technology Facilities Council, UK; Ministry of Science and Higher
Education, Poland.

Part of this work was carried out on the Dutch national
e-infrastructure with the support of the SURF Cooperative through
grant e-infra 160022 \& 160152. The LOFAR software and dedicated
reduction packages on \url{https://github.com/apmechev/GRID_LRT} were
deployed on the e-infrastructure by the LOFAR e-infragroup, consisting
of J.\ B.\ R.\ Oonk (ASTRON \& Leiden Observatory), A.\ P.\ Mechev (Leiden
Observatory) and T. Shimwell (ASTRON) with support from N.\ Danezi
(SURFsara) and C.\ Schrijvers (SURFsara). This research has made use of the University
of Hertfordshire high-performance computing facility
(\url{https://uhhpc.herts.ac.uk/}) and the LOFAR-UK compute facility,
located at the University of Hertfordshire and supported by STFC
[ST/P000096/1]. The J\"ulich LOFAR Long Term Archive and the German
LOFAR network are both coordinated and operated by the J\"ulich
Supercomputing Centre (JSC), and computing resources on the
supercomputer JUWELS at JSC were provided by the Gauss Centre for
supercomputing e.V. (grant CHTB00) through the John von Neumann
Institute for Computing (NIC). This research made use of {\sc Astropy}, a
community-developed core Python package for astronomy
\citep{AstropyCollaboration13} hosted at
\url{http://www.astropy.org/}, and of {\sc Matplotlib} \citep{Hunter07}.


\section*{Data Availability}

The sample of AGN used in this paper will be released with the published version of the paper.


\bibliographystyle{mnras}
\bibliography{references}


\appendix


\bsp
\label{lastpage}
\end{document}